\documentclass[]{aastex631}

\usepackage{bm}
\newcommand\Vec{\bm}

\shorttitle{Acoustic wave properties in footpoints of coronal loops}
\shortauthors{Riedl et al.}

\begin{document}

\title{Acoustic wave properties in footpoints of coronal loops in 3D MHD simulations}

\correspondingauthor{Tom Van Doorsselaere}
\email{tom.vandoorsselaere@kuleuven.be}

\author[0000-0002-2327-3381]{Julia M. Riedl}
\affiliation{Centre for mathematical Plasma Astrophysics, Department of Mathematics, KU Leuven, Celestijnenlaan 200B bus 2400, 3001 Leuven, Belgium}

\author[0000-0001-9628-4113]{Tom Van Doorsselaere}
\affiliation{Centre for mathematical Plasma Astrophysics, Department of Mathematics, KU Leuven, Celestijnenlaan 200B bus 2400, 3001 Leuven, Belgium}

\author[0000-0002-1820-4824]{Fabio Reale}
\affiliation{Dipartimento di Fisica \& Chimica, Universitá di Palermo, Piazza del Parlamento, 90134, Palermo, Italy}

\author[0000-0002-7830-7147]{Marcel Goossens}
\affiliation{Centre for mathematical Plasma Astrophysics, Department of Mathematics, KU Leuven, Celestijnenlaan 200B bus 2400, 3001 Leuven, Belgium}

\author[0000-0002-9882-1020]{Antonino Petralia}
\affiliation{INAF-Osservatorio Astronomico di Palermo, Piazza Parlamento 1, 90134 Palermo, Italy}

\author[0000-0001-5274-515X]{Paolo Pagano}
\affiliation{Dipartimento di Fisica \& Chimica, Universitá di Palermo, Piazza del Parlamento, 90134, Palermo, Italy}



\begin{abstract}
Acoustic waves excited in the photosphere and below might play an integral part in the heating of the solar chromosphere and corona. However, it is yet not fully clear how much of the initially acoustic wave flux reaches the corona and in what form. We investigate the wave propagation, damping, transmission, and conversion in the lower layers of the solar atmosphere using 3D numerical MHD simulations. A model of a gravitationally stratified expanding straight coronal loop, stretching from photosphere to photosphere, is perturbed at one footpoint by an acoustic driver with a period of 370 seconds. For this period acoustic cutoff regions are present below the transition region (TR). About 2\% of the initial energy from the driver reach the corona. The shape of the cutoff regions and the height of the TR show a highly dynamic behavior. Taking only the driven waves into account, the waves have a propagating nature below and above the cutoff region, but are standing and evanescent within the cutoff region. Studying the driven waves together with the background motions in the model reveals standing waves between the cutoff region and the TR. These standing waves cause an oscillation of the TR height. In addition, fast or leaky sausage body-like waves might have been excited close to the base of the loop. These waves then possibly convert to fast or leaky sausage surface-like waves at the top of the main cutoff region, followed by a conversion to slow sausage body-like waves around the TR.
\end{abstract}



\section{Introduction}

The solar atmosphere perpetually exhibits all kinds of oscillations and waves, may it be in the corona \citep{tomczyk_etal_2007}, in the chromosphere \citep{morton_etal_2012}, or in the photosphere in the form of \textit{p}-modes \citep{leighton_1960} or in sunspots \citep{khomenko_collados_2015_review}. Studying and understanding waves in the solar atmosphere is crucial for both coronal seismology, where the plasma properties of the corona are inferred from the wave behavior (see the reviews by \citealt{nakariakov_verwichte_2005_review} and \citealt{nakariakov_kolotkov_2020_review}), and to explain coronal and chromospheric heating (see the reviews by \citealt{vandoorsselaere_etal_2020_review} and \citealt{srivastava_etal_2021_review}). Although waves can be generated by impulsive events like flares \citep{liu_etal_2011, zhang_etal_2020} or in the atmosphere itself by a steady background flow \citep{nakariakov_etal_2016, karampelas_van_doorsselaere_2021}, it is believed that most waves stem from convective motions in or below the photosphere, including the internal \textit{p}-modes. 

Typical wave periods observed in the corona peak at around three to five minutes \citep{de_moortel_etal_2002, van_doorsselaere_etal_2008_b, tomczyk_mcintosh_2009, morton_etal_2016, morton_etal_2019}, which suggests a connection to the photospheric \textit{p}-modes with the same peak period. However, waves excited by \textit{p}-modes have predominantly an acoustic character and low frequency waves with periods around five minutes are subjected to an acoustic cutoff layer present in the chromosphere. In this layer, acoustic waves with frequencies below the cutoff frequency are evanescent and non-propagating (standing). It is thus questioned how \textit{p}-mode excited waves of typical coronal periods propagate into higher layers of the solar atmosphere.

Several mechanisms have been proposed. The best established one is the decrease of the effective cutoff frequency for slow (magneto)acoustic waves that propagate along an inclined magnetic field \citep[and many more]{bel_leroy_1977, de_pontieu_etal_2004, de_pontieu_etal_2005, mcintosh+jefferies_2006, jefferies_etal_2006, rajaguru_etal_2019}. The effective cutoff frequency can also be reduced by radiative losses, if the radiative relaxation time is small enough \citep{centeno_etal_2006, khomenko_etal_2008_letter, felipe+sangeetha_2020}. However, according to \citet{heggland_etal_2011} changes in the radiative relaxation time, which characterizes radiative losses through Newton's cooling law, have little effect, when more sophisticated treatments for radiative losses are applied. 

Furthermore, around the equipartition layer, where the Alfv\'{e}n speed equals the sound speed, mode conversion from acoustic waves excited by \textit{p}-modes to low $\beta$ fast magnetoacoustic and Alfv\'{e}n waves can take place \citep{schunker_cally_2006, cally+goossens_2008, cally+hansen_2011}. If that conversion from waves of predominantly acoustic to predominantly magnetic properties occurs in time, the waves are not affected by the acoustic cutoff region. The converted fast waves can then scatter in Fourier space and manifest as kink waves \citep{cally_2017, cally_khomenko_2019_part_I}. That \textit{p}-modes can convert to tube waves (sausage, kink, or fluting modes) was also shown by \citet{bogdan_etal_1996}, \citet{hindman_jain_2008}, \citet{gascoyne_etal_2014}, and \citet{riedl_etal_2019}. Observationally, there is vast evidence of tube waves in the solar atmosphere \citep{verwichte_etal_2005, morton_etal_2011, nistico_etal_2013, anfinogentov_etal_2015, grant_etal_2015, frij_etal_2016, keys_etal_2018, gilchrist_etal_2021}.

Considering the wave-altering effects mentioned above, namely the frequency dependent acoustic cutoff region, magnetic field inclination, radiative effects, and mode conversion - which are further dependent on the atmospheric and magnetic structure - considerably complicates the study of waves through the highly stratified lower solar atmosphere, where all of these effects interplay. In addition, fast wave refraction \citep{khomenko_collados_2006} and wave reflection from the transition region (TR) have to be taken into account. Reflection of waves between the TR and the temperature minimum could lead to the formation of a chromospheric acoustic resonator \citep{zhugzhda+locans_1981, botha_etal_2011,felipe_2019,jess_etal_2020}, which further influences the wave behavior. To understand these complex processes and their interaction we have to rely on numerical simulations, that do not only provide much higher resolution than observations, but also allow us to isolate specific phenomena by assuming certain simplified models. 

The numerical simulations conducted by \citet{heggland_etal_2011} were already very realistic in the sense that a sophisticated treatment of radiative losses was included and that the waves were driven self-consistently by convective motions. However, their analysis focused on the bigger picture and they did not study single wave trains and their conversion in particular. \citet{khomenko+cally_2012} study the conversion of acoustic waves to Alfv\'{e}n waves in sunspots at the equipartition layer, while \citet{santamaria_etal_2015} investigate mode conversion and reflection in a magnetic arcade with a potential magnetic field. They include not only an equipartition layer, but also a cutoff layer, a null point, and the TR in their model. A similar study is done by \citet{tarr_etal_2017}. \citet{khomenko_cally_2019_part_II} verify the theoretical work of \citet{cally_2017} and show that the amount of flux that reaches the corona is increased when a flux tube structure is present. The simulations of \citet{riedl_etal_2019} also employ a model with straight tightly packed flux structures, that includes a TR. There they show that, depending on the field inclination, initially planar acoustic waves convert to waves with magnetic properties, and that both sausage and kink waves are excited. However, a cutoff layer was excluded by utilizing a high frequency driver. The cutoff frequency is numerically studied by \citet{felipe+sangeetha_2020} for different parameters like magnetic field strength and inclination, radiative losses, and different hydrostatic (HS) equilibria. However, they use a plane-parallel model and constant magnetic field, which is both a strong simplification.  Simulations of waves in a stratified coronal loop of constant radius including TR regions were conducted by \citet{belien_etal_1999}, where an Alfv\'{e}n wave driver was used. The waves eventually converted to low $\beta$ slow magnetoacousic waves that caused density perturbations, however, this was not studied in conjunction with an acoustic cutoff layer.

In this paper we utilize three-dimensional (3D) magnetohydrodynamic (MHD) simulations to further investigate the wave behavior of initially plane acoustic waves in the lower solar atmosphere. In particular, we look at the wave propagation, damping, conversion/transformation, interaction, and transmission through the steep TR into the corona in the footpoint of a coronal loop, that is locally driven by a monochromatic gravity-acoustic wave with a period of $T=370$ s at the base of the photosphere. The whole simulation domain contains a full gravitationally stratified loop that stretches from photosphere to corona to photosphere and significantly expands in the chromospheres and the corona. At the footpoints of the loop, a $\beta=1$ layer is present, as well as an acoustic cutoff region for the driver period. The paper is organized as follows. In Section \ref{sec:5_methods} we explain the model used for the simulations (Section \ref{subsec:5_model}), the numerical setup (Section \ref{subsec:5_numerics}) and how we calculated the locations of the acoustic cutoff region (Section \ref{subsec:5_cutoff_region}). We present and discuss our results in the following three sections, where we quantify the damping and energy transmission into the corona (Section \ref{sec:5_energy_trans_into_corona}), investigate the interaction of the waves with the highly dynamic background, including strong changes in the height of the TR (Section \ref{sec:5_interaction_of_driver_waves_with_background}), and determine the excited wave modes through comparison with simplified theoretical models (Section \ref{sec:5_determination_of_wave_modes}). Finally, we present our conclusions in Section \ref{sec:5_conclusions}.

\section{Methods} \label{sec:5_methods}
\subsection{Model} \label{subsec:5_model}


As an initial condition for our simulations, we use the model of \citet{reale_etal_2016}, featuring a straight 3D loop spanning from photosphere to photosphere in a cylindrical coordinate system. 
While the plasma parameters at the footpoints do not exactly correspond to photospheric values but to conditions a few hundred kilometers above, the footpoints are still located below the temperature minimum. Thus, for the remainder of this study, we regard the bottom and top of the computational domain as the photosphere.
The loop is tenuous and relatively cool and thus appropriate for quiet Sun conditions. The model was obtained following the procedure used by \citet{guarrasi_etal_2014}: A HS equilibrium atmosphere with straight vertical magnetic field and increased plasma pressure and magnetic field strength in the center of the loop is numerically relaxed by conducting a preliminary 2.5D simulation (a 2-dimensional simulation but using three vector components for the velocity and the magnetic field). The initial HS equilibrium already includes a chromosphere, a TR, and a corona. The chromosphere is isothermal and works as a mass reservoir. During the equilibration process the loop significantly expands in the corona to balance the horizontal pressure disparity. As a last step, the 2.5D model is spanned to an axisymmetric 3D model. 

We use a gravitational profile of
\begin{equation}
    g(z)=-g_\odot \cos{\left(\frac{(z-z_0)\pi}{L}\right)},
\end{equation}
where $g_\odot=274$ m s$^{-2}$ is the gravitational acceleration at the solar surface, $z_0$ is the z-coordinate of the photosphere at the lower boundary of our domain, and $L=62.61$ Mm is the total length of the loop. This results in the gravitational acceleration pointing to the negative $z$-direction for $z<0$ and to the positive $z$-direction for $z>0$ with $g(0)=0$ m s$^{-2}$ at the loop apex. Thus, the gravity along our straightened loop is the same as for a curved loop. Gravity across the loop, as well as the loop curvature, are neglected. 

\begin{figure}
  \centering
  \includegraphics[width=0.7\textwidth]{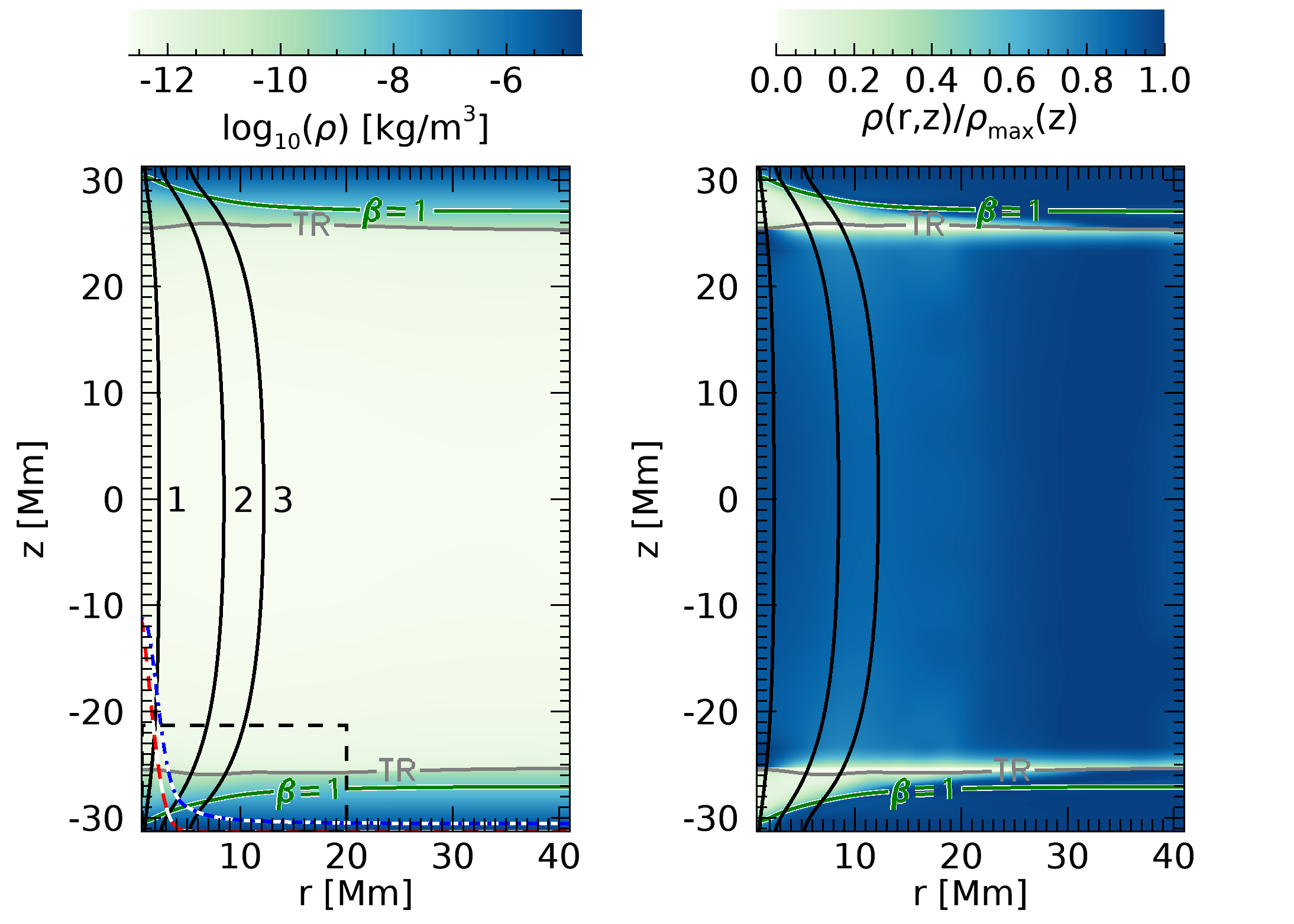}
  \caption{\textit{Left:} Logarithm of the density in a vertical cut through the model atmosphere. The TRs and $\beta=1$ layers are shown as horizontal gray and green lines, respectively. The vertical black lines with numbers highlight three magnetic field lines. The magnitude of the total magnetic field at the bottom boundary is indicated by the blue dash-dotted line, while the driver amplitude is shown as red dashed line (both in arbitrary units). The section marked with dashed black lines shows the plot range of later figures. \textit{Right:} Horizontal density structure shown by the relative density with respect to the maximum density for all $z$.}
  \label{fig:5_full_domain}%
\end{figure}

Figure \ref{fig:5_full_domain} (left) shows the axisymmetric density profile of the model along the loop. The TR and $\beta=1$ contours are indicated by the gray and green lines, respectively. For simplicity, the TR is defined as the location where the temperature equals  $40~000$ K. This location is very close to the location of the steepest temperature gradient, i.e. the strongest downwards thermal conduction flux. Since the TR has a finite thickness, all TR lines should be regarded as approximate positions.

While the vertical stratification of the model due to gravitational effects is evident, there are also horizontal structures present. The horizontal structuring can be appreciated in the right panel of Figure \ref{fig:5_full_domain}, which shows the normalized density as a function of the $r$-coordinate for all heights $z$, i.e. the relative horizontal density profiles at every $z$ location. It is normalized with respect to the maximum density at every height. The loop is located at the left side of the plots in Figure \ref{fig:5_full_domain}, with the loop axis technically being located at $r=0$ Mm. However, the region around $r=0$ Mm is not included in the simulation domain to avoid a singularity. The total magnetic field across the loop is approximately Gaussian-shaped and is shown for the bottom of the domain by the blue dash-dotted line. At the footpoints, the magnetic field reaches a maximum of 273 G and decreases to 13 G at the loop apex at $z=0$ Mm. The expansion of the loop is shown by the three magnetic field lines marked with numbers 1 to 3. Field line 2 is rooted at the full-width half-maximum (FWHM) of the total magnetic field at the footpoint (blue dash-dotted line) and acts as a proxy to define the loop radius. Field line 1 is thus located inside the loop, whereas field line 3 is located outside.

\subsection{Numerical setup and boundary conditions} \label{subsec:5_numerics}

To reduce computing time, only one quarter of the loop cylinder is simulated, which is possible due to the symmetry of our system. Our simulation domain ranges from $0.73$ Mm to $41.01$ Mm in the $r$-direction, from $0$ to $\pi/2$ in the $\phi$-direction, and from $-31.31$ Mm to $31.31$ Mm in the $z$-direction, with $192\times128\times768$ data points. In the inner quarter of the grid in the $r$-direction, the resolution is uniform with 76 km per cell. For higher $r$, the grid is stretched, leading to linearly decreasing resolution until the outer boundary with a maximum cell size of 1500 km. In the $z$-direction, starting from the upper and lower boundary, the grid is homogeneous with the highest resolution of 30 km per cell until after the highly stratified TR. In the much less stratified corona, the grid is stretched with lowest resolution in the middle of the domain (568 km per cell) at the loop apex. In the vertical direction, the resolution corresponds to the resolution of the simulations of \citet{reale_etal_2016}, while the resolution of our model in the horizontal directions is twice as coarse as the original model with the cell sizes doubled and the number of cells halved.

The simulations are performed using the \textsc{Pluto} code \citep{mignone_etal_2007,mignone_etal_2012,mignone_etal_2018}. The MHD equations are solved in 3D cylindrical coordinates using the Harten-Lax-Van Leer approximate Riemann solver with a piecewise total variation diminishing (TVD) linear reconstruction method for the spatial integration. For time advancement a second order Runge-Kutta scheme is used with Strang operator splitting \citep{strang_1968}. Gravity is included using a vector body force and the div$B=0$ constraint is controlled by the eight wave formulation described by \citet{powell_etal_1999}. Thermal conduction along the magnetic field is included and applied using the super-time-stepping technique \citep{alexiades_etal_1996}. Radiative losses and external heating rates are not taken into account. For simplicity, we also assume full ionization throughout the whole domain. 

We utilize reflective boundary conditions for both boundaries in the $r$-direction and periodic boundaries in the $\phi$-direction. Following \citet{reale_etal_2016}, we incorporate a reflective boundary condition but with reversed sign for the tangential components of the magnetic field for the upper $z$ boundary. At the lower boundary, the same boundary conditions are set, however, for the vertical velocity, the density, and the pressure a perturbation according to an analytical solution for a vertical gravity-acoustic wave \citep{mihalas_mihalas_1984,khomenko+cally_2012,santamaria_etal_2015} is added
\begin{eqnarray}
    v_z'&=&A \exp \left(\frac{z}{2H}+\Im(k_z)z\right) \sin \left( \omega t - \Re(k_z)z\right), \label{eq:driver_velocity}\\
    p' &=& A p |P| \exp \left(\frac{z}{2H}+\Im(k_z)z\right) \sin \left( \omega t - \Re(k_z)z + \phi_P\right), \label{eq:driver_pressure}\\
    \rho' &=& A \rho |R| \exp \left(\frac{z}{2H}+\Im(k_z)z\right) \sin \left( \omega t - \Re(k_z)z + \phi_R\right). \label{eq:driver_density}
\end{eqnarray}
Here, $A=300$ m s$^{-1}$ is the amplitude, $p|P|$ and $\rho|R|$ are the relative amplitudes for the pressure and density perturbation, respectively, $H$ is the pressure scale height, $k_z$ is the vertical wave number, which only has a real part in our case, $\omega=2\pi/T$ with period $T=370$ s is the driver frequency, and $\phi_P$ and $\phi_R$ are the pressure and density phase shift compared to the velocity perturbation. The details of the driver are explained by \citet{santamaria_etal_2015}. The chosen amplitude is consistent with velocities measured in the photosphere \citep[e.g.][]{beck_etal_2009}. This period is chosen as it is close to the typical 5-minute period of \textit{p}-modes while at the same time leading to an acoustic wave frequency that has a cutoff layer in the chromosphere (see Section \ref{subsec:5_cutoff_region}). 

The driver is applied locally at one footpoint in the interior of the loop, which is achieved by multiplying Equations \ref{eq:driver_velocity} to \ref{eq:driver_density} with the Gaussian
\begin{equation} \label{eq:5_gaussian}
    G(r)=\exp \left( -\frac{r^2}{ \sigma^2} \right), ~~ \sigma=2 ~ \mathrm{Mm}.
\end{equation}
The function $G(r)$ is plotted as red dashed line in Figure \ref{fig:5_full_domain} (left). The figure also shows that while the driver amplitude is still high for field line 1, it is already much lower for field line 2 and nearly zero for field line 3. 

Even though the model is in a relaxed state, it still contains background velocities of up to 18 km s$^{-1}$. Below the TR, the background velocities still have values in the same order as the driver amplitude. Therefore, in order to distinguish effects of driven waves from background motions, we also conduct simulations without a driver for comparison. This allows us to extract the perturbed quantities by subtracting the quantities from the simulation without a driver from the quantities of the simulation with driver, as was also done by \citet{riedl_etal_2021}. For simulations without a driver the bottom boundary conditions are the same as the boundary conditions at the top of the domain.

The main interest of this study lies only within the small sub-domain shown by the dashed box in Figure \ref{fig:5_full_domain} (left). Even so, for stability reasons the whole domain is simulated. Waves propagating away from the loop are eventually damped and no significant wave energy flux reaches the reflective outer boundary. The upper boundary at $z=31.31$ Mm and the outer boundary at $r=41.01$ Mm are sufficiently far away from the driver location for reflected waves from these boundaries to be of no concern.

We note that for the symmetry at hand of both model and driver, a 2.5D simulation would have been sufficient to obtain our results. However, we still conducted 3D simulations in order to facilitate comparisons to future simulations that break the symmetry.




\subsection{Determination of the cutoff region} \label{subsec:5_cutoff_region}

In stratified atmospheres acoustic waves with frequencies below the cutoff frequency are not allowed to propagate. For 5-minute acoustic waves as excited by \textit{p}-modes, such a cutoff layer is present in the chromosphere. For a non-magnetic isothermal atmosphere, \citet{lamb_1909} derived the cutoff frequency to be 
\begin{equation} \label{eq:5_wc_lamb}
    \omega_\mathrm{c}=\frac{v_\mathrm{s}}{2H},
\end{equation}
where $v_\mathrm{s}$ is the speed of sound and $H$ is the pressure scale height. In an isothermal atmosphere, $\omega_\mathrm{c}$, as calculated by Equation \ref{eq:5_wc_lamb}, is constant. However, for a more general atmosphere the local acoustic cutoff frequency can be estimated by using the local values of $v_\mathrm{s}$ and $H$. 

The pressure scale height $H$ is a measure of how fast the atmosphere changes with height. If in a plane-parallel HS equilibrium atmosphere an acoustic wave propagates in an oblique direction to the gravitational stratification, for example along an inclined magnetic field, the scale height is effectively increased for the wave, which results in a decreased cutoff frequency of
\begin{equation} \label{eq:5_wc_lamb+ramp}
    \omega_\mathrm{c}=\frac{v_\mathrm{s}}{2H}\cos\left(\theta\right),
\end{equation}
where $\theta$ is the inclination angle of the magnetic field with respect to the direction of gravitational acceleration \citep{bel_leroy_1977,de_pontieu_etal_2004,de_pontieu_etal_2005,jefferies_etal_2006,mcintosh+jefferies_2006}. This decrease of the effective cutoff frequency is called the ``ramp effect'' or ``magnetoacoustic portal'' .

Since our model is not only vertically stratified but also includes variations in the horizontal direction, a simple multiplication of a factor $\cos\left(\theta\right)$ is not enough to include the effect of the magnetic field direction. Instead, we use the following definition of the scale height
\begin{equation}
    H=-\frac{p}{dp/d[l]},
\end{equation}
where $dp/d[l]$ is either the derivative of the pressure in vertical direction, which effectively is a localized version of Equation \ref{eq:5_wc_lamb}, or $dp/d[l]$ is the pressure derivative in the magnetic field direction, which corresponds to Equation \ref{eq:5_wc_lamb+ramp}. Thus, in the following the cutoff frequency for acoustic waves is determined by
\begin{eqnarray}
    \omega_{\mathrm{c},z}&=\frac{v_\mathrm{s}}{2p}\left|\frac{\partial p}{\partial z}\right| ~~~ & \mathrm{vertical~wave~propagation}, \label{eq:5_omega_c_z}\\
    \omega_\mathrm{c,\parallel}&=\frac{v_\mathrm{s}}{2p}\left|\frac{\partial p}{\partial s}\right| ~~~ & \mathrm{field\text{-}aligned~wave~propagation}, \label{eq:5_omega_c_para}
\end{eqnarray}
with $s$ the coordinate along the considered field line. 

\begin{figure}
    \centering
    \includegraphics[width=0.4\textwidth]{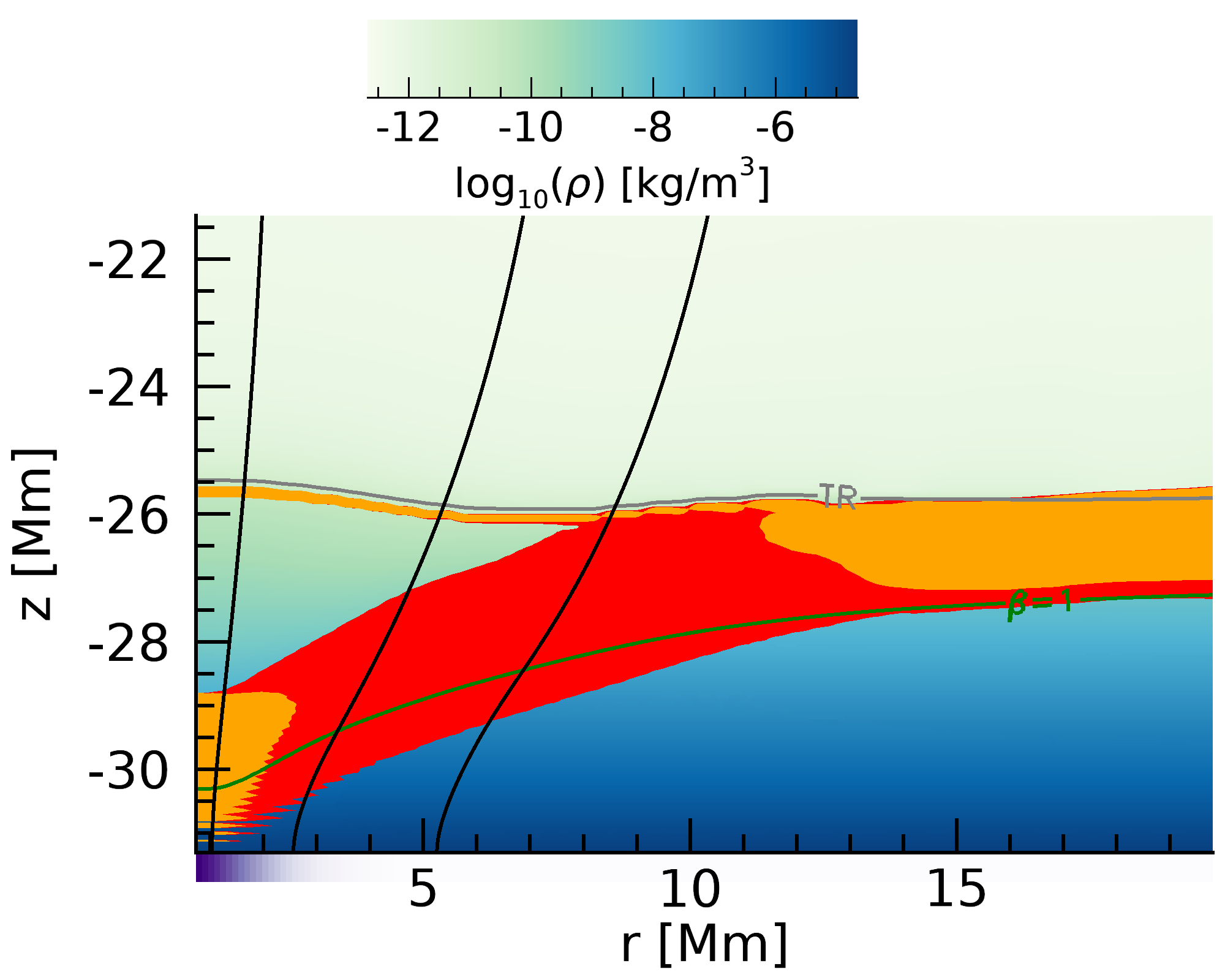}
    \caption{Acoustic cutoff regions for vertical (red, Equation \ref{eq:5_omega_c_z}) and field-aligned wave propagation (orange, Equation \ref{eq:5_omega_c_para}) for waves with a period of $T=370$ s. The background color shows the logarithm of the density with the same color scale as in Figure \ref{fig:5_full_domain} (left). The TR and $\beta=1$ layer are marked by the gray and green line, respectively. Field lines 1 to 3 are shown as black lines. The purple bar underneath the $x$-axis shows the strength of the driver.}
    \label{fig:5_cutoff_region}
\end{figure}

Figure \ref{fig:5_cutoff_region} shows the resulting cutoff regions for acoustic waves with frequencies equal to the driver frequency, i.e. the regions where the cutoff frequency is higher than the driver frequency. The plot range corresponds to the dashed box in Figure \ref{fig:5_full_domain} (left). The cutoff region for the field-aligned wave propagation (orange region) is less strict than for the vertical wave propagation (red region), leading to the orange region being generally a subset of the red region, i.e. all orange regions are also included in red regions. It is immediately apparent that there is no extensive cutoff layer for field-aligned acoustic waves at the border of the loop around field line 2 or outside the loop around field line 3. However, such a layer is present for the inner part of the loop, where the driver is strongest. The strength of the driver, as modified by the Gaussian in Equation \ref{eq:5_gaussian}, is indicated by the purple bar below the plot.

It is important to note that the values for the cutoff frequency calculated by Equations \ref{eq:5_omega_c_z} and \ref{eq:5_omega_c_para} are only an estimate. \citet{felipe+sangeetha_2020} did a linear parameter study with different HS equilibrium model atmospheres and magnetic field strengths and inclinations and found significant differences between various analytical formulas of the cutoff frequency and between these formulas and the numerically determined cutoff frequency. They conclude that using more sophisticated formulas like those presented in \citet{deubner_gough_1984}, \citet{schmitz_fleck_1998}, and \citet{roberts_2006} do not lead to a significant improvement of the cutoff frequency over the formula derived by \citet{lamb_1909}, which is used as a basis in this study. In fact, the cutoff regions determined by the formulas mentioned in the three papers above, with the derivatives and scale heights adapted as for Equations \ref{eq:5_omega_c_z} and \ref{eq:5_omega_c_para}, are very similar to the cutoff regions shown in Figure \ref{fig:5_cutoff_region}. The main difference is a thicker second cutoff region around the TR \citep[especially for the formula by][]{roberts_2006} and a cutoff layer being already present at the bottom of the domain for the formula by \citet{deubner_gough_1984}. The second cutoff region around the TR is subjected to some uncertainties for all formulas, because the numerical derivatives are not well behaved due to the sudden changes of the gradients.


%

\section{Energy transmitted into the corona} \label{sec:5_energy_trans_into_corona}

The simulations are carried out for a duration of four driver periods. The wave energy flux is calculated in the same way as in \citet{riedl_etal_2021} with
\begin{equation} \label{eq:5_wave_flux}
      \Vec{F}=-\frac{1}{\mu_0}\left( \Vec{v}\times\Vec{B}\right)\times\Vec{B} + \left( \frac{\rho v^2}{2}+\rho \Phi + \frac{\gamma}{\gamma-1}p\right) \Vec{v},
\end{equation}
where we insert the perturbed quantities determined by subtracting the quantities of the simulation without driver from the ones of the simulations with driver. This allows us to distinguish the wave energy flux from the total, mostly oscillatory energy flux, with the latter having much higher amplitudes \citep[Sect. 4]{bogdan_etal_2003}. The first term of Equation \ref{eq:5_wave_flux} is the Poynting flux $\Vec{S}$, whereas the other terms describe the hydrodynamic (HD) part of the flux $\Vec{F}_\mathrm{HD}$. We stress that removing the time-dependent background is not equivalent to a linearization and all non-linear effects should be retained. However, while our results clearly show the presence of non-linearities, we do not focus on them in detail. Some non-linear effects are discussed in the next section (Section \ref{sec:5_interaction_of_driver_waves_with_background}).

Similar to the analysis in \citet{riedl_etal_2019}, we split the vector components into components parallel to the magnetic field and normal to the magnetic flux surfaces. The vector components in azimuthal direction, i.e. components in the $\phi$-direction of the cylindrical domain, are negligible due to the symmetry of the system and therefore ignored. 

\begin{figure}
    \centering
    \includegraphics[width=0.45\textwidth]{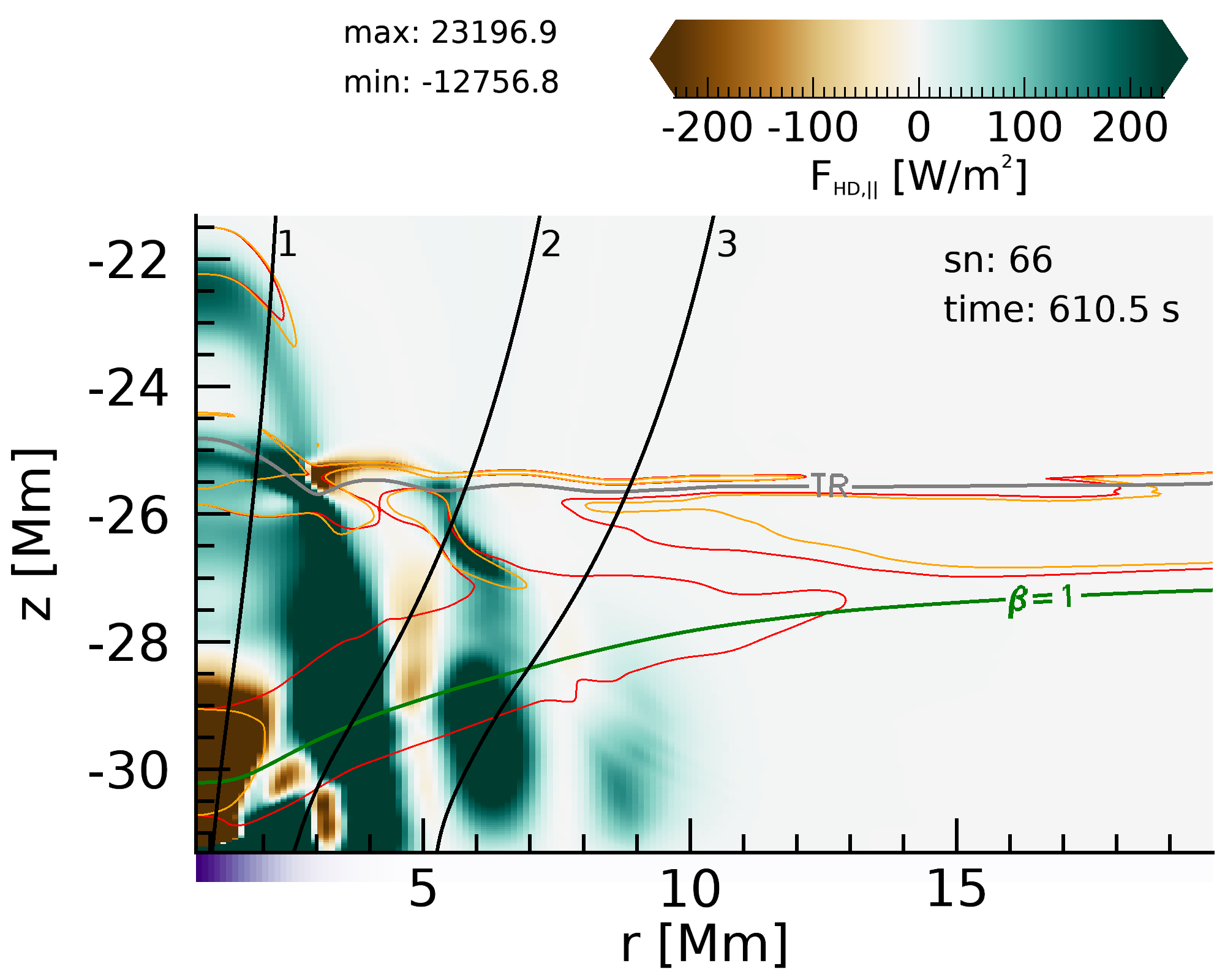}
    \caption{Hydrodynamic flux component parallel to the magnetic field in a selected snapshot of the time sequence. The snapshot number and simulation time is given in the top right. The color scale is saturated. The actual value range for the entire time sequence of four periods for the shown plot range is printed next to the color bar. The cutoff regions according to Equations \ref{eq:5_omega_c_z} and \ref{eq:5_omega_c_para} are given as red and orange contours, respectively. The TR and $\beta=1$ layer are indicated by the gray and green lines, respectively. Field lines 1 to 3 are drawn as black lines. The strength of the driver is indicated by the purple bars below the $x$-axis of the plots. The full time sequence is available as an animation online and covers four driver periods. The video duration is 8 seconds.}
    \label{fig:5_parallel_HD_flux_time_series}
\end{figure}

Due to the nature of the driver, most energy flux lies within the HD component of the flux parallel to the magnetic field. Within the considered subset of the domain marked in Figure \ref{fig:5_full_domain} (left), $|F_\mathrm{HD,\parallel}|\approx 10^3 \times |F_\mathrm{HD,\perp}|\approx 10^1-10^2 \times |S_\parallel|\approx 10^5 \times |S_\perp|$.
A selected snapshot (number 66) of the time sequence of $F_\mathrm{HD,\parallel}$ is shown in Figure \ref{fig:5_parallel_HD_flux_time_series}. The whole time sequence is available as a movie in the online version of the article. Even with a cutoff region present, wave flux propagates up and reaches the TR. However, the flux of the initial wave is not transmitted into the corona (around snapshot 53, $t=490$ s). This could either be a hint to a thicker second cutoff region around the TR as predicted by the analytic formulas of \citet{deubner_gough_1984}, \citet{schmitz_fleck_1998}, and \citet{roberts_2006}, or it occurs because the reflection due to the large gradients is too strong. The wave could also lose its energy to a surface wave excited at and along the steep TR \citep{erdelyi_etal_2007,fedun_etal_2011}. However, the next wave of positive parallel HD flux is transmitted (snapshot 66, $t=611$ s, displayed in Figure \ref{fig:5_parallel_HD_flux_time_series}). This also occurs at later times for snapshot 99 ($t=916$ s) and around snapshot 137 ($t=1267$ s). This clearly shows that at least some of the acoustic wave flux reaches the corona. The $\beta=1$ layer seems to have little influence, as there is no significant amount of Poynting flux generated. This is not entirely unexpected for a vertical loop, especially if the wave vector aligns with the magnetic field direction.

When studying the time sequence of Figure \ref{fig:5_parallel_HD_flux_time_series} strong changes  of the cutoff region shapes are visible. These changes occur partly due to the presence of background motions in the imperfectly relaxed model, but also partly due to the influence of the driver. The difference between these effects is discussed in the next section. The elongated horizontal cutoff structure in the first few snapshots of the time sequence occurs due to the background motions and corresponds to a single initial wave propagating up as soon as the simulation starts. The change of the shapes of the cutoff regions due to the driver can be regarded as non-linear effect, because the wave amplitudes are strong enough to change the plasma parameters significantly, which results in a change of the local cutoff frequency. The role of the dynamics of the cutoff regions on the wave damping is very difficult to determine, because several effects play a role at the same time. However, it is plausible that if the dynamic behavior is strong enough, the cutoff frequency may be locally decreased enough during short time frames to open up for acoustic waves of certain frequencies.

Strikingly, there are many waves present outside the loop beyond the driven location, which are visible as vertical green shapes for $r\approx 5$ Mm and larger. These are laterally leaked waves that were also present in the two-dimensional (2D) simulations of \citet{riedl_etal_2021}, which clearly shows that lateral leakage can also be expected around the footpoints of coronal loops in a 3D configuration. The waves leak out of the loop in both the $\beta>1$ and $\beta<1$ region, which validates the results from \citet{riedl_etal_2021} that where limited by having a $\beta>1$ plasma in the whole domain. We can therefore conclude that lateral wave leakage is to be expected around any flux concentration in the lower solar atmosphere.

\begin{figure}
    \centering
    \includegraphics[width=0.5\textwidth]{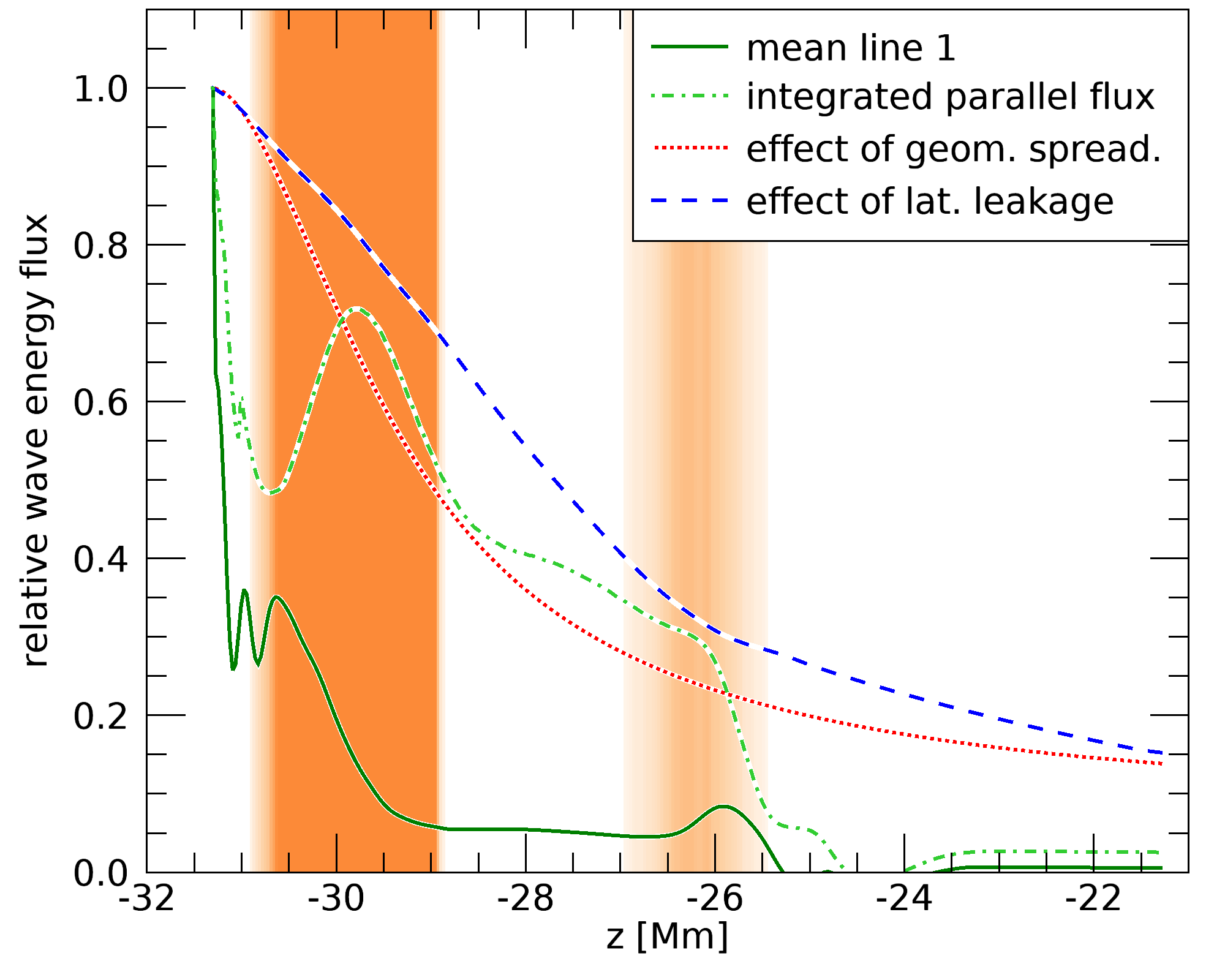}
    \caption{Relative wave energy flux parallel to the magnetic field as a function of height for simulation times from $t=3T$ to $t=4T$. The solid dark green line shows the time-averaged energy flux along field line 1. The dash-dotted light green line shows the flux integrated over time and integrated over the area inside the loop until field line 2 (energy within loop). The dotted red line shows the $R^{-2}$ line for field line 2, whereas the dashed blue line shows the expected damping due to lateral leakage for field line 3. The background color shows the time-dependent location of the cutoff region for field-aligned wave propagation (where $\omega<\omega_\mathrm{c,\parallel}$) at field line 1. The stronger the color, the more snapshots include a cutoff region at a specific $z$-location.}
    \label{fig:5_flux_damping}
\end{figure}

To quantify the wave energy flux damping and to determine which fraction of energy reaches the corona we repeat the analysis also done by \citet{riedl_etal_2021} with the present simulation data. Figure \ref{fig:5_flux_damping} shows the relative wave energy flux as a function of height for simulation data spanning one driver period $T$ at the end of the time sequence, i.e. from $t=3T$ to $t=4T$. For these times the laterally leaked flux has already reached the location of field line 3 for all heights considered in our analysis, leading to periodic wave behavior at this location. Therefore, as opposed to the analysis by \citet{riedl_etal_2021}, it is not necessary to define different time-integration and averaging boundaries for different vertical coordinates $z$. 

The solid dark green line in Figure \ref{fig:5_flux_damping} shows the total (Poynting and HD) mean parallel flux for field line 1, averaged in the given time-frame. The dotted red line shows the effect of damping expected due to geometric spreading caused by the expansion of the loop, which is equal to $R(z)^{-2}$ with $R(z)$ the distance from the loop axis to field line 2. In \citet{riedl_etal_2021}, a similar line was drawn in their Figure 8 but for $R(z)^{-1}$ due to the 2D setup. The blue dashed line in Figure \ref{fig:5_flux_damping} is an estimate for the damping caused by lateral flux leakage, which is dependent on the ratio between the total flux lost laterally at the considered field line and the total incoming flux. It is determined as described in \citet{riedl_etal_2021} for the location of field line 3, with the additional (3rd) dimension accounted for by integrating along areas instead of lines. The driver amplitude at the location of the root of field line 3 at the bottom of the domain has already decreased by a factor of $10^{-3}$. Therefore, field line 3 can be considered to be located outside the driven region. 

\citet{riedl_etal_2021} have given an estimate of how to convert the effects of lateral leakage in the 2D setup to the expected effect lateral leakage has in 3D. Unfortunately, we cannot test the validity of that estimate by a comparison to the current simulations, because of the strong differences in the model and the driver frequency. However, considering similar height ranges, we can say that the damping due to lateral wave leakage is far weaker for the simulations of this study than for those in \citet{riedl_etal_2021}.

Even below the main cutoff region, which is indicated by the first orange shade in the background, the mean energy flux of field line 1 (solid dark green line) is damped much stronger than expected for the combined effects of geometric spreading and lateral wave leakage (red dotted line times blue dashed line). At the deepest point below the main cutoff region, the relative flux has decreased to about 26\%, whereas the combined geometric damping effects would only explain a decrease to about 96\%. 
We find a significant amount of reflected flux below the main cutoff region, as well as below the TR. These reflections can at least partly explain the strong damping of the flux below the main cutoff region, however, we were unable to quantify the strength of this effect.
Interestingly, this strong initial damping exactly coincides with the damping observed by \citet{gilchrist_etal_2021} in their solar pore 3. However, given that our model is quite different from a solar pore with a much weaker magnetic field, the similarities could be a coincidence. 


In the cutoff region itself the flux is strongly damped, as expected. This suggests that the actual cutoff region in our simulations corresponds with the cutoff region determined by Equations  \ref{eq:5_omega_c_z} or \ref{eq:5_omega_c_para}. Between $z=-26.5$ Mm and $z=-24$ Mm the mean energy flux shows a strange behavior, which is connected to the dynamically moving TR and the interaction between driven waves and background motions. These effects are discussed in Section \ref{sec:5_interaction_of_driver_waves_with_background}. Above the TR in the corona, the combined effects of geometric spreading and lateral wave leakage (red dotted line times blue dashed line) estimate the parallel flux on a single field line to have decreased to about 2\% or less of the initial value, whereas the real damping obtained from our simulations for field line 1 (solid dark green line) shows a decrease down to 0.5\% or less.

To estimate how much energy from the driver is transmitted into the corona, we integrate the parallel wave energy flux over the entire loop cross section, with the radius defined by field line 2, and over one driver period from $t=3T$ to $t=4T$. This results in the total wave energy as a function of height inside the loop. The relative energy is shown in Figure \ref{fig:5_flux_damping} by the dash-dotted light green line. After initially strong damping the energy increases at the location of the cutoff region of field line 1. This occurs because of the lateral leakage of waves, which in turn may occur due to the refraction of the waves caused by the sound speed profile. The sound speed has a minimum roughly following the $\beta=1$ layer and the cutoff region for vertically propagating waves (regions enclosed by red contours in Figure \ref{fig:5_parallel_HD_flux_time_series}). The escaping wave fronts in Figure \ref{fig:5_parallel_HD_flux_time_series} propagate along this sound speed minimum. The maximum of parallel flux outside the cutoff region for field-aligned wave propagation (orange contours) but inside the loop (left of field line 2) is located at the same height as the energy peak in Figure \ref{fig:5_flux_damping} around $z\approx -30$ Mm. Due to the oblique propagation of these leaking waves, the ``blobs'' of parallel flux stay longer at similar heights, which has an impact for the time integration. 

At the TR above $z \approx -26$ Mm the energy again drops significantly and only a small amount of flux is transmitted through. In the corona, 10 Mm above the footpoint of the loop, only 2\% of the initial energy remains.
From the location just below the main cutoff region at the local minimum of the light green energy line, which is just after the strong initial damping of the dark green curve, about 4\% of the energy is transmitted into the corona.
From the chromosphere, about 5\% of the energy is transmitted into the corona. 
This is higher than the transmission value from the chromosphere of around 1\% found by the non-magnetic simulations of \citet{erdelyi_etal_2007} using a 300 s driver. 

The sharp drop of transmitted energy through the TR could not only be accounted for by wave reflection due to the strong gradients, but could also occur because the waves excite and lose energy to surface waves that horizontally propagate along the sharp TR \citep{erdelyi_etal_2007, fedun_etal_2011}. Although the time sequence of the TR movement supports this theory (see gray line in the animation of Figure \ref{fig:5_parallel_HD_flux_time_series}), we do not investigate this effect further in the current work.

\section{Interaction of driven waves with background motions} \label{sec:5_interaction_of_driver_waves_with_background}

\begin{figure}
    \centering
    \includegraphics[width=0.5\textwidth]{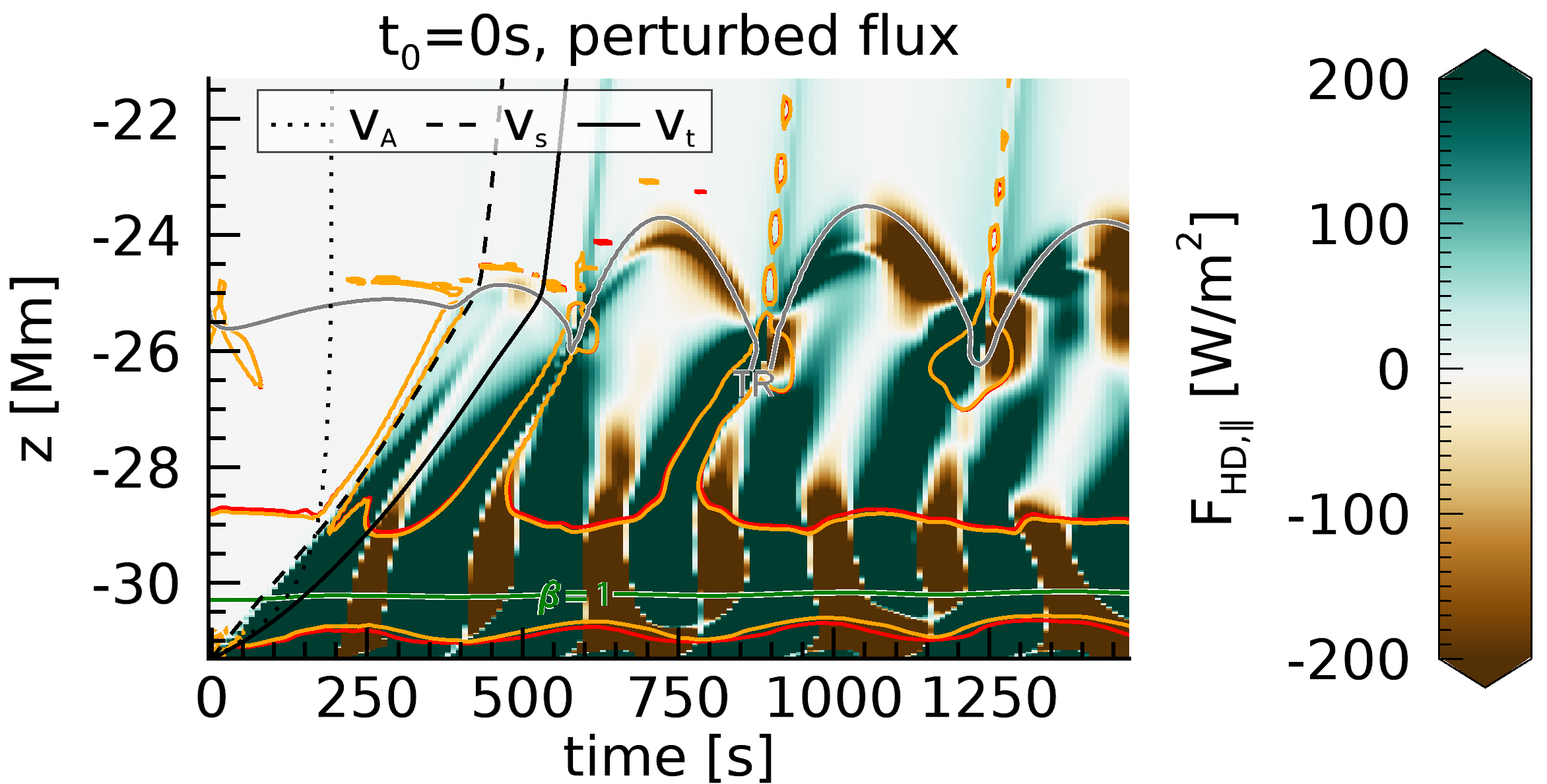}\includegraphics[width=0.5\textwidth]{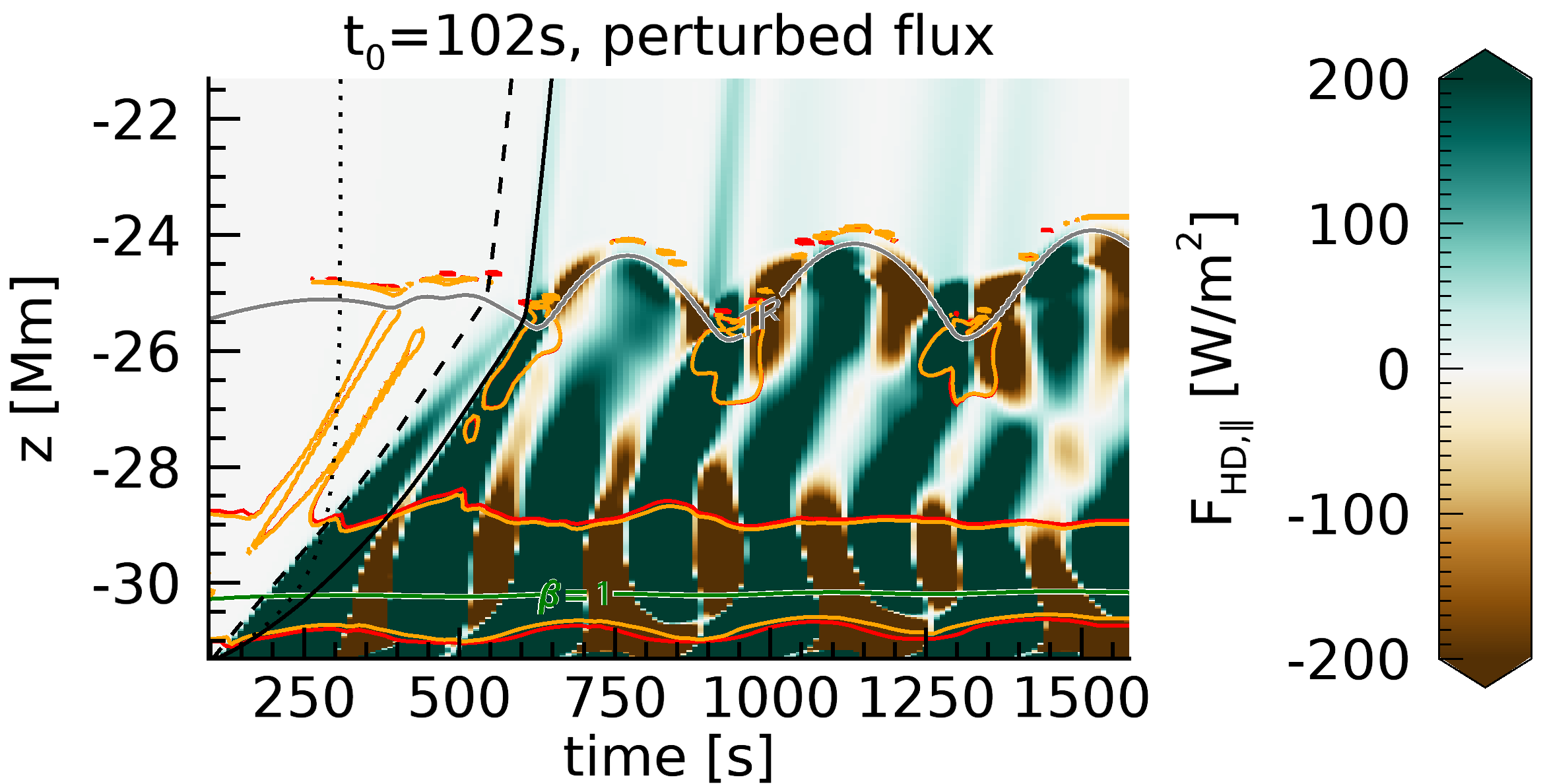}
    \includegraphics[width=0.5\textwidth]{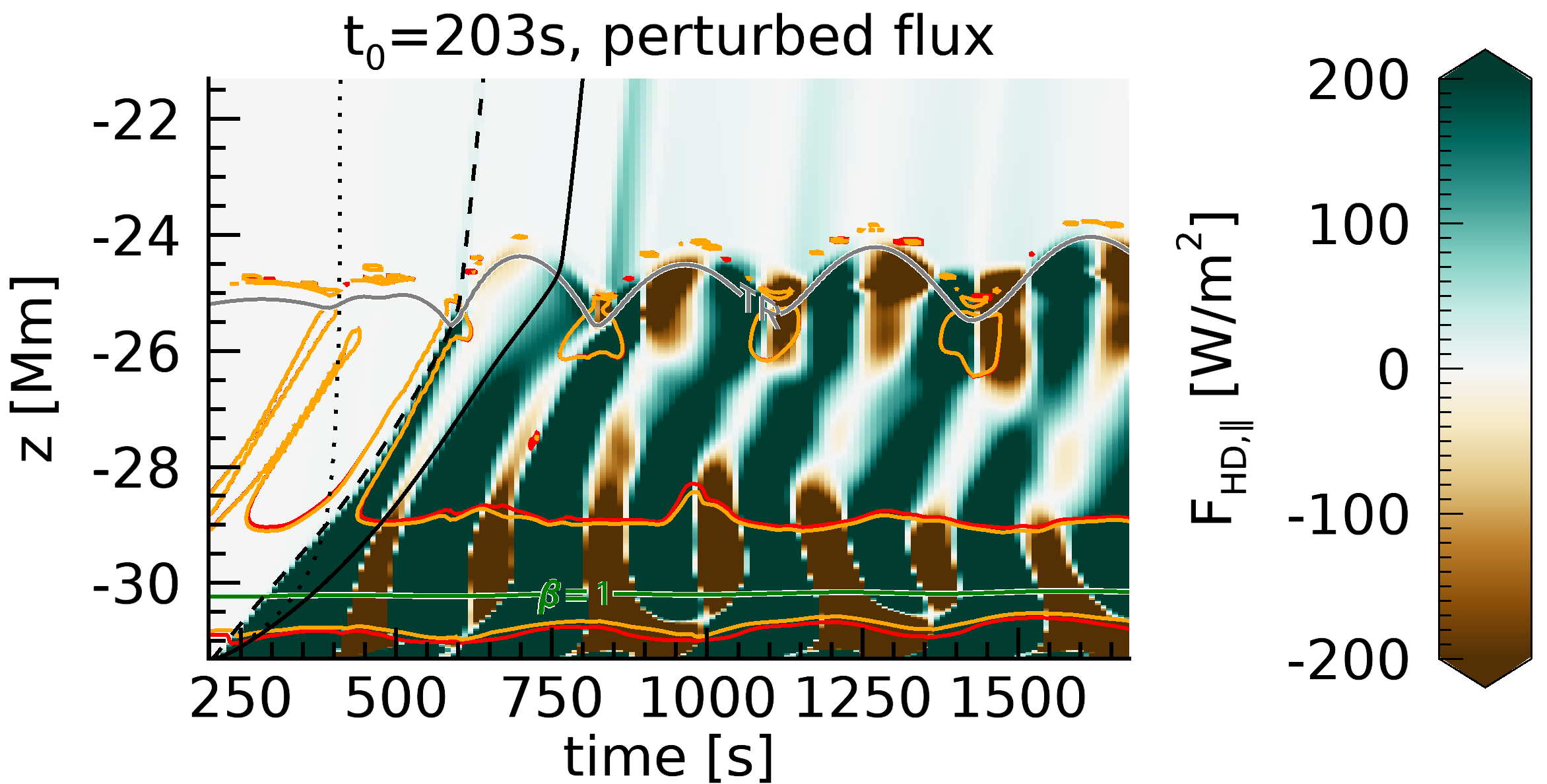}\includegraphics[width=0.5\textwidth]{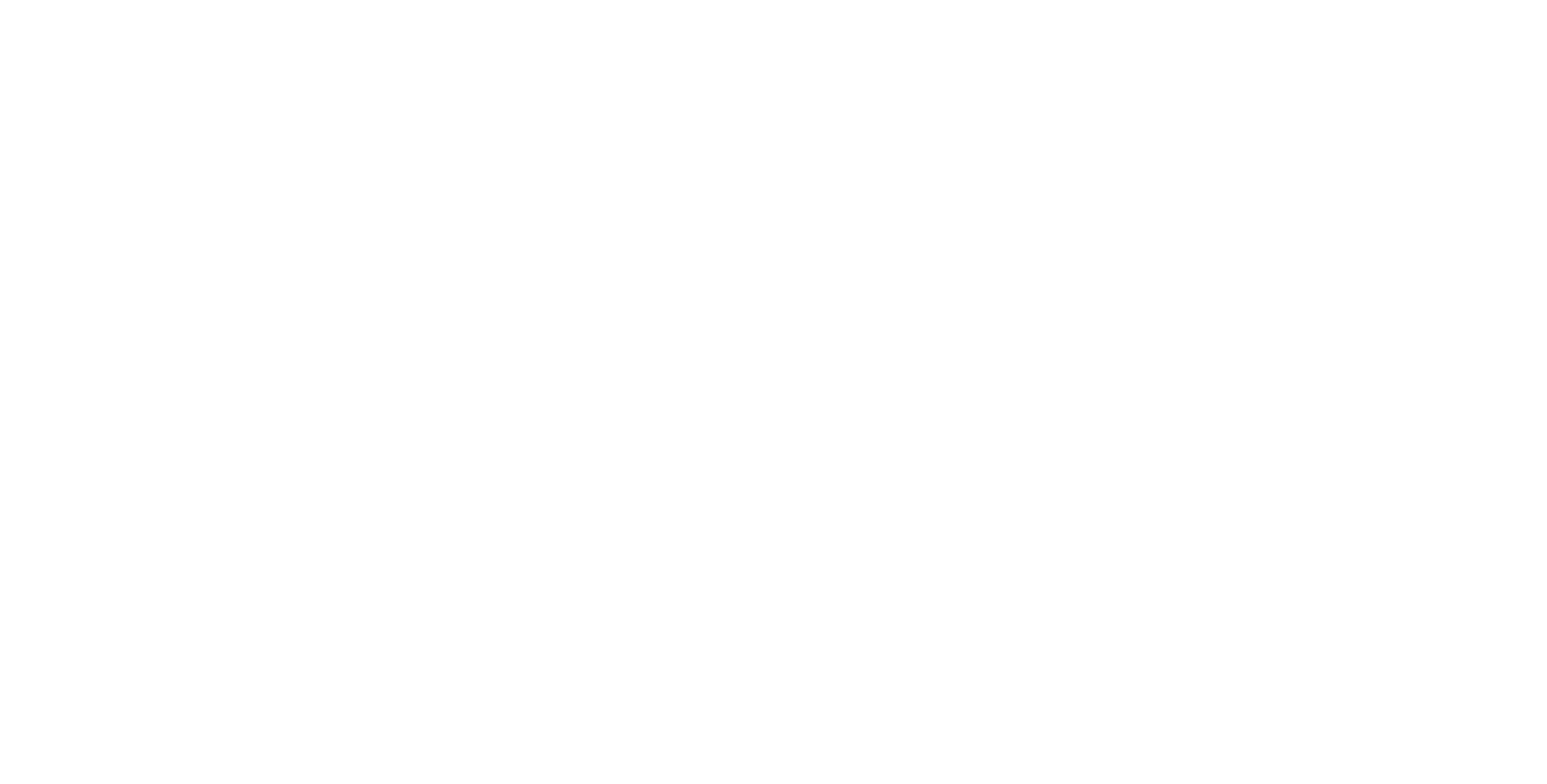}
    \vspace{0.3cm}
    
    \includegraphics[width=0.5\textwidth]{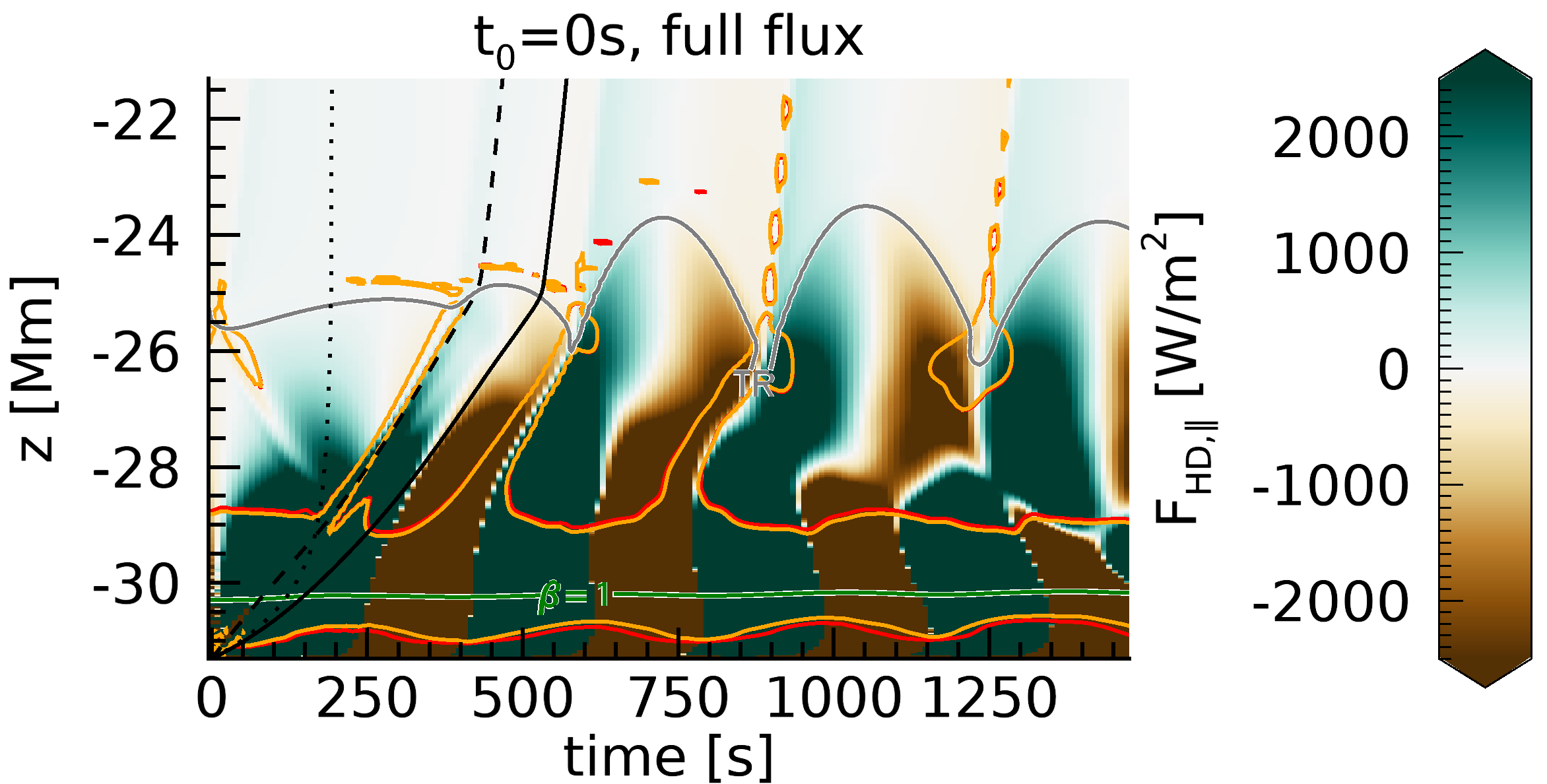}\includegraphics[width=0.5\textwidth]{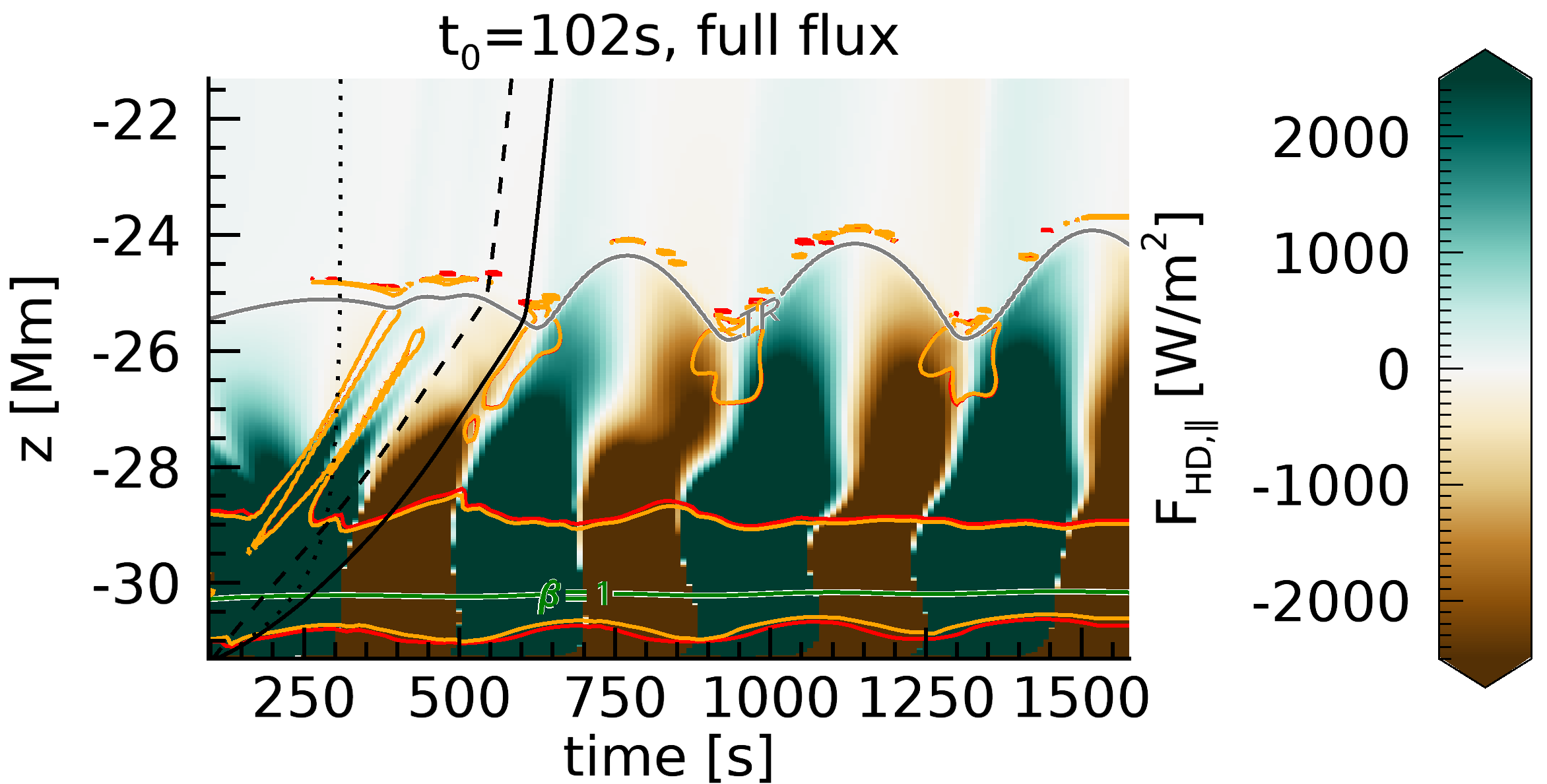}
    \includegraphics[width=0.5\textwidth]{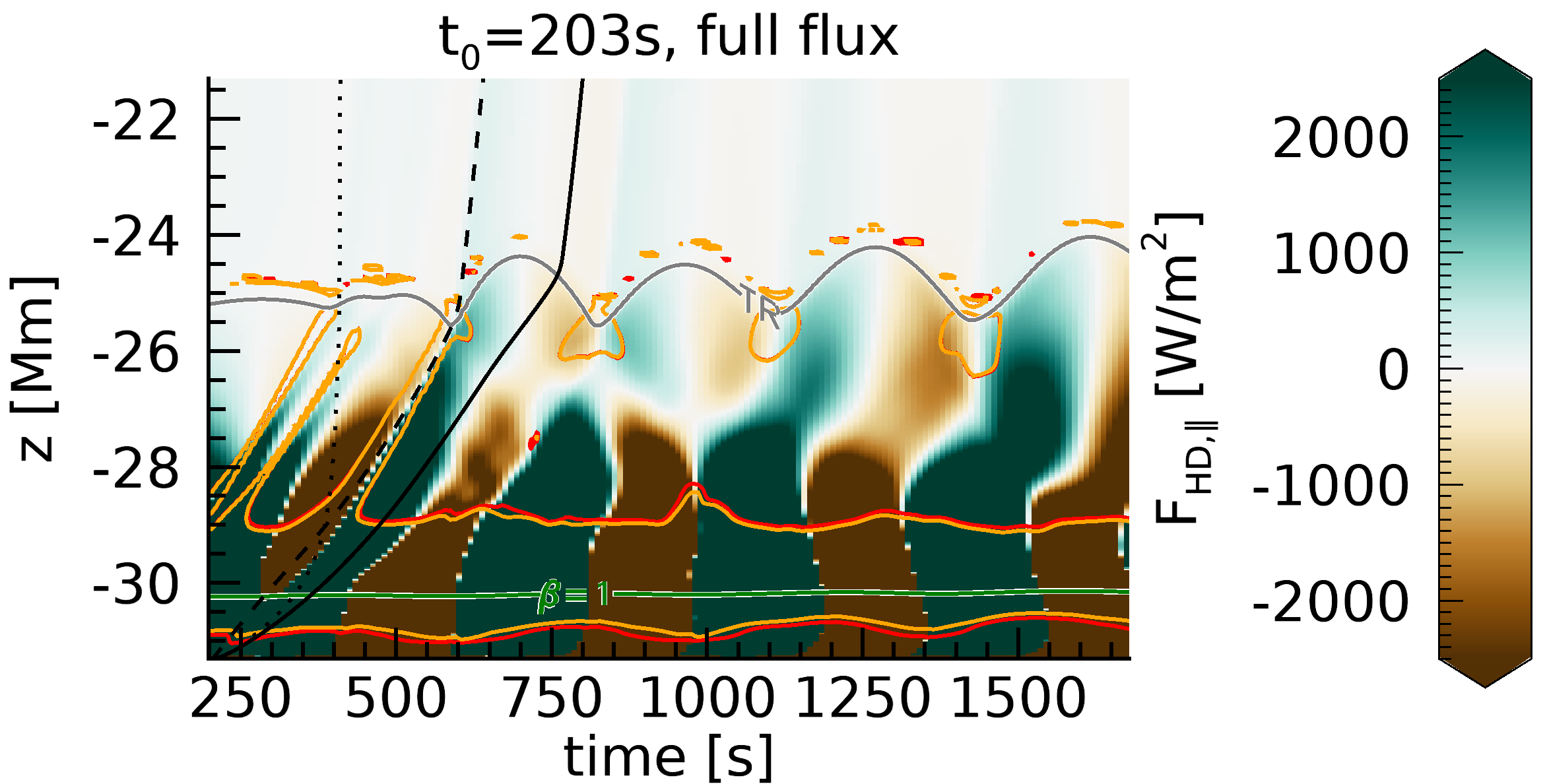}\includegraphics[width=0.5\textwidth]{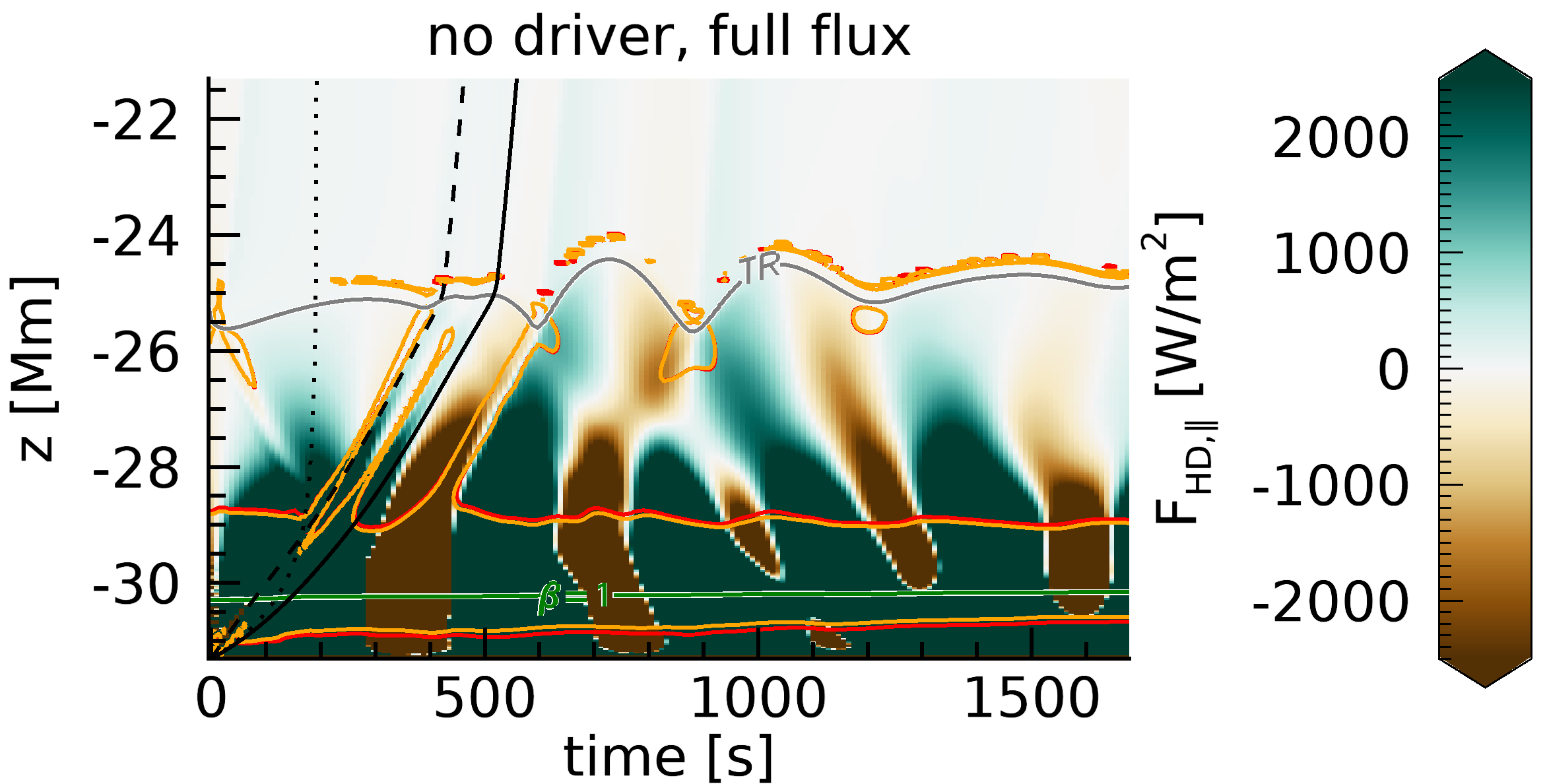}
    \caption{Height-time diagrams of the HD energy flux component parallel to the magnetic field along field line 1 for all conducted simulations. The top three panels show the wave energy flux calculated using perturbed quantities, while the bottom four panels show the full energy flux. The black lines show the local Alfv\'{e}n speed (dotted), sound speed (dashed), and tube speed (solid). The cutoff regions are indicated by the red (vertical propagation) and orange (field-aligned propagation) contours. The TR and $\beta=1$ layer are indicated by the gray and green lines, respectively. The color scale is saturated. The bottom right panel covers a longer time range than the other panels.}
    \label{fig:5_HT_graphs}
\end{figure}

The top left panel of Figure \ref{fig:5_HT_graphs} shows the height-time diagram for the parallel HD wave energy flux component along field line 1. The initial wave propagates with approximately the sound speed (dashed black line), as is expected from acoustic waves. Below and above the cutoff region the wave fronts are inclined from the vertical i.e. the waves are propagating. Inside the cutoff region between $z\approx -31$ Mm and $z\approx -29$ Mm, however, standing waves (i.e. vertical patterns) are formed as is expected by the theory of linear acoustic waves in an isothermal HS atmosphere. According to this theory, the wavenumber is purely imaginary for vertical waves with frequencies below the acoustic cutoff frequency, which means that only standing waves can exist \citep{priest_2014}. This shows that the predicted cutoff region from Equations \ref{eq:5_omega_c_z} and \ref{eq:5_omega_c_para} coincide very well with the actual cutoff region in that area. 


Above $z=-29$ Mm and below the TR (gray line) the highly variable cutoff region shows elongated shapes, as separated parts of the cutoff region (visible for example in Figure \ref{fig:5_parallel_HD_flux_time_series} at $z\approx -26$ Mm and $z\approx -22$ Mm) propagate upward. The positive parallel flux (green shapes) shows different inclinations close to those elongated cutoff regions, and is therefore propagating with different speeds. The third wave train at $t\approx 550$ s is so steep that it is nearly a standing wave even between the cutoff regions. Possibly, the real cutoff frequency in that area is not very well captured by the theoretical value.

It is obvious from the animation of Figure \ref{fig:5_parallel_HD_flux_time_series} and from Figure \ref{fig:5_HT_graphs} that not only the cutoff regions but also the height of the TR strongly depends on time. These changes are only partly caused or modified by the non-linear effects of the waves excited by the driver and are also present in the simulation without driver, as can be seen in the bottom right panel of Figure \ref{fig:5_HT_graphs}\footnote{The time range of this panel is different to the other panels to include the full simulation times of all panels.}, which shows the energy flux of the background motions. While this highly dynamic background state is usually not desired in controlled numerical simulations, it can be regarded as a feature in this case. In the real solar atmosphere, there are always some kind of plasma motions present. Especially the location of the TR is highly variable in reality. Other simulations \citep{heggland_etal_2011} have shown similarly strong dynamics of the cutoff region and TR height.

However, there is definitely an influence of the driver on the cutoff regions and especially the TR. In the simulation with driver (first panel of Figure \ref{fig:5_HT_graphs}) the bottom of the cutoff region shows more variability, as well as additional cutoff regions at greater heights. While the time of the peaks of the TR height approximately matches between the simulations with and without driver, the TR movement covers a wider height-range for the driven simulation. Acoustic flux is mainly transmitted into the corona at the points of lowest TR heights. 

To distinguish the effects of the background motions from the driven waves on the dynamic development of the atmosphere, we conducted two additional simulations that are identical to the previous simulation with driver except that the driver is switched on at later times, namely at $t_0=102$ s (second panel of Figure \ref{fig:5_HT_graphs}) and $t_0=203$ s (third panel of Figure \ref{fig:5_HT_graphs}). Therefore, the driven waves of all simulations propagate through different background atmospheres. The simulations with $t_0\neq0$ s show very similar features to the simulation with $t_0=0$ s, except that the propagating waves above the main cutoff region are more regular. In addition, the TR height changes are less prevalent for the simulations with $t_0\neq0$ s. The stronger change in the TR height for the $t_0=0$ s simulation can be explained by the first driven waves propagating up with the same speed at the same time as the background motions arising at the beginning of the simulation due to imperfect boundary conditions. The background motions and the driver therefore act constructively for pushing the TR.

While the TR height also shows a somewhat periodic behavior for the $t_0\neq0$ s simulations, the periodicity is different and the peaks occur at different times. This is especially true if the simulations are run for a longer time. We tested this by running the simulation with $t_0=102$ s and the simulation without driver for another four periods. For the additional run time, the TR height and cutoff region shape displays the same nearly mono-periodic behavior for the simulation with $t_0=102$ s as can be seen in the top right panel of Figure \ref{fig:5_HT_graphs}. There is no mono-periodic behavior for the simulation without driver. The most striking difference for the time ranges plotted in Figure \ref{fig:5_HT_graphs} can be seen for the simulation with $t_0=203$ s (third panel ). There, the last TR valley occurs at the same time ($t\approx 1400$ s) as the last hill for the simulation without driver (last panel of Figure \ref{fig:5_HT_graphs}). Similarly, the last hill for the driven simulation occurs at the same time ($t\approx 1600$ s) as the last valley for the simulation without driver. This shows that the driven waves clearly have an influence on the dynamic behavior of the atmosphere, which arises due to their complex interaction with the background motions. 

All three driven simulations show a formation of standing waves just below the TR. The standing waves are connected to the TR movement with positive flux where the TR moves up and negative flux where it moves down. Other than the cutoff region contours shown in the plots, that are based on the isothermal cutoff frequency found by \citet{lamb_1909}, the theoretical cutoff frequencies based on the formulas by \citet{deubner_gough_1984}, \citet{schmitz_fleck_1998}, and \citet{roberts_2006} all predict a second cutoff region around the TR for the whole duration of the simulations. The wave behavior in the simulations agrees with this second cutoff region. 

Between the top of the main cutoff region ($z\approx -29$ Mm) and the lowest height of the TR ($z\approx-26$ Mm) the waves in the driven simulations are clearly propagating up, as the wave trains have a positive finite slope. However, when not focusing on the wave energy flux obtained by inserting only perturbed quantities ($q'=q_\mathrm{driver}-q_\mathrm{nodriver}$) into Equation \ref{eq:5_wave_flux}, and instead taking the full energy flux into account (inserting $q=q_\mathrm{driver}$ or $q=q_\mathrm{nodriver}$, respectively, into Equation \ref{eq:5_wave_flux}), the same region shows mainly standing waves (bottom four panels of Figure \ref{fig:5_HT_graphs}). In other words, the driven waves itself are mainly propagating, but the interaction of these waves with the overall motions lead to standing waves. We assume this to be the case due to reflections from the TR of both the background motions (see patterns with negative inclination in the last panel of Figure \ref{fig:5_HT_graphs}) and some part of the driven waves. Reflections of the driven waves from the TR at the considered height range cannot be appreciated in the top three panels of Figure \ref{fig:5_HT_graphs}, however, they may simply not be well captured by Equation \ref{eq:5_wave_flux} when the background is removed. The superposition of the upward and downward propagating waves then leads to the formation of (nearly) standing waves. This behavior has certain similarities with a chromospheric resonator \citep{zhugzhda+locans_1981, botha_etal_2011,felipe_2019,jess_etal_2020}. There is also the possibility that these waves are not standing at all but propagating with the very high Alfv\'{e}n speed (dotted line) in that region. However, with the time resolution of our snapshots this is impossible to tell.


The period of the waves in the panels showing the perturbed flux (top three panels of Figure \ref{fig:5_HT_graphs}) is about half the driver period. On the other hand, the panels showing the full flux of the driven waves (bottom panels of Figure \ref{fig:5_HT_graphs} with the exception of the bottom right panel) show the same periodicity as the driver. This is not due to some complicated non-linear interaction, but due to the definition of $F_\mathrm{HD,\parallel}$. For the full flux, the term in the right bracket of Equation \ref{eq:5_wave_flux} is always positive, while $v_\parallel$ oscillates around 0 with the driver period. However, for the perturbed flux, both the term in the bracket and $v_\parallel$ oscillate around 0 with the driver period but with a phase shift between them. When these terms are multiplied with each other, the resulting frequency of the perturbed flux is higher than the driver frequency. Due to the same reason, namely the definition of $F_\mathrm{HD,\parallel}$, the full flux peaks at different locations above the TR than the perturbed flux. However, here the full flux mainly shows oscillatory behavior, while the perturbed flux is purely positive and thus corresponds to an increase in energy.

The standing waves visible in the full energy flux for the driven simulations roughly correspond to the standing waves just below the TR visible in the perturbed energy flux. They have exactly the same periodicity as the TR movement. Therefore, it is likely that the complex interaction between driven waves and background motions in turn drives the TR oscillations. Since the TR peaks for the driven simulations do not occur at the same time as for the simulation without driver, it can be assumed that the change of the TR height is driven by the standing waves and not vice versa. 

The period of the standing waves and TR movement are slightly different for all simulations, however, for the given simulation time they all increase with time. The elapsed time between TR valleys for the simulation with $t_0=203$ s equals 200 s, 240 s, 290 s, and 310 s, while for the simulation without driver it is 200 s, 300 s, 320 s, and 410 s. These periods are mostly shorter than the driver period of $T=370$ s. For the increased run time of the simulation with $t_0=102$ s (8 driver periods instead of 4), the periodicity of the TR movement settles at around 380 s, which is close to the driver frequency. The TR oscillations in the simulation without driver do not show any mono-periodicity even with the extended run time.

\begin{figure}
\centering
\includegraphics[width=0.9\textwidth]{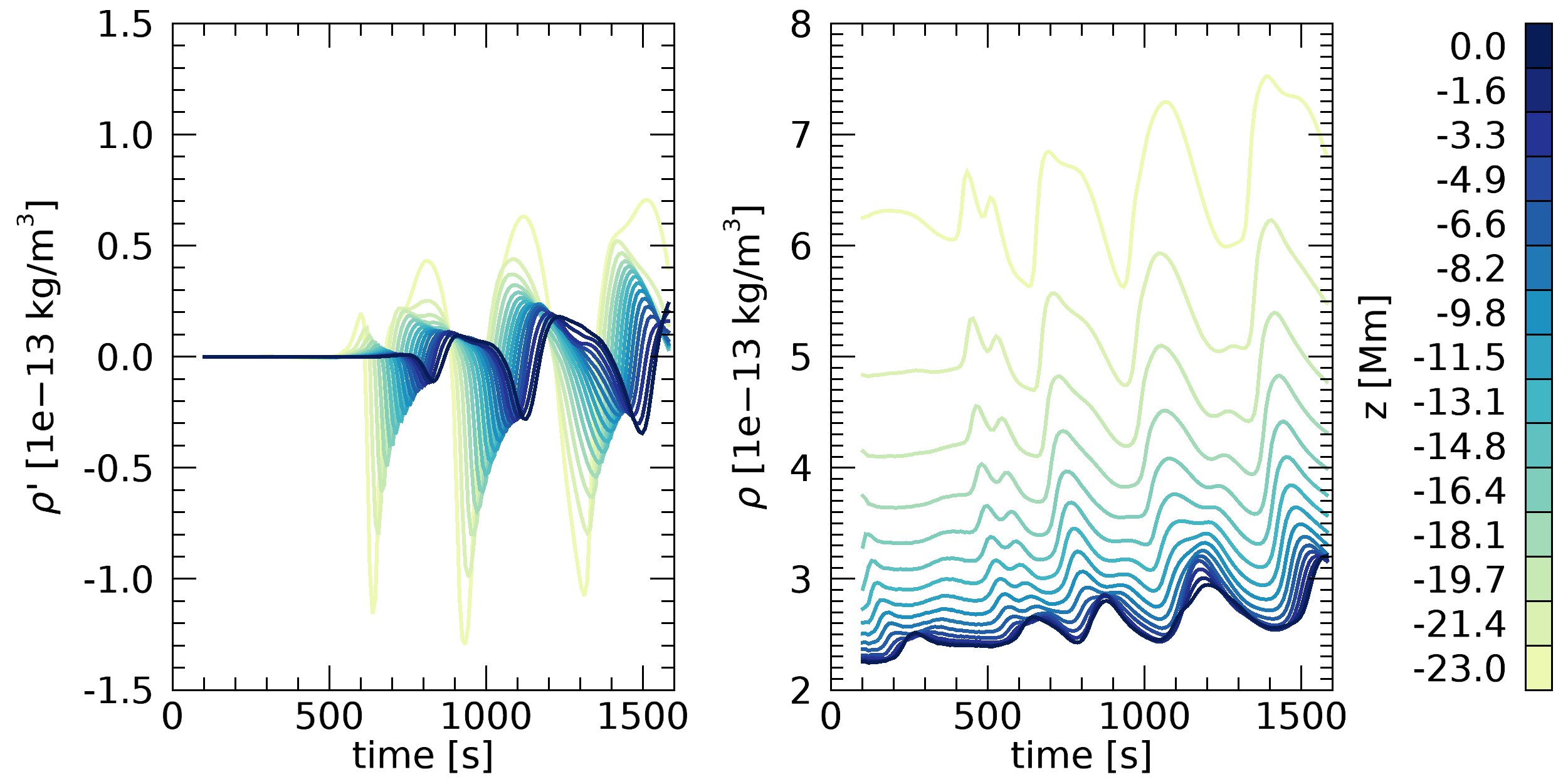}
\caption{Density perturbation (left) and density (right) as a function of time for different heights (shown in different colors) above the TR for the simulation with $t_0=102$ s at the location of field line 1. The color bar indicates the $z$-location of the different lines. We note that the values of subsequent lines in the right panel decrease because the density decreases with height.}
\label{fig:5_denisty_profiles_with_height}
\end{figure}

It is of interest to ascertain how the waves reaching the corona would be detected observationally. The most straight-forward way for detecting waves is by observing the intensity, which is proportional to the density squared. The left panel of Figure \ref{fig:5_denisty_profiles_with_height} shows the time series of the density perturbation for field line 1 between the TR and the loop apex for the simulation with $t_0=102$ s. The waves do not show sinusoidal shapes for any heights. However, when looking at the full density (right panel), which determines the observed intensity, the waves appear more sinusoidal, especially close to the apex of the loop. This appearance was validated by looking at the power spectrum of the waves. The more or less sinusoidal shape for the full density agrees with the observations. On the other hand, sinusoidal-like waves also appear in our simulations without driver, albeit without alteration by the component of the perturbed density displayed in the left panel of Figure \ref{fig:5_denisty_profiles_with_height}. This is not entirely unexpected because also the TR oscillates for the simulations without driver. Nevertheless, the driver alters the wave behavior. The increasing trend in the right panel of Figure \ref{fig:5_denisty_profiles_with_height} shows a net flux of mass towards the loop apex, which was also the case for \citet{belien_etal_1999}.

\section{Determination of wave modes} \label{sec:5_determination_of_wave_modes}

Due to the symmetry of the setup, all waves in our simulations are automatically axisymmetric. In the following, we determine whether these waves convert from plane waves to tube waves in the cylindrical structure, like in \citet{riedl_etal_2019}, and which wave modes are excited, or if the original plane nature of the waves remains. For the symmetry at hand, the only possible tube waves are sausage waves. 

We investigate the possible sausage wave behavior by comparing our simulations with theoretical calculations of sausage waves in a homogeneous cylinder embedded in a homogeneous background. The magnetic field inside and outside the cylinder is parallel to the cylinder axis and gravity is neglected. While this simple idealized model with a radial step-function dependence certainly has its shortcomings, exactly this model is also used to identify tube wave modes for observational data \citep[to name a few]{morton_etal_2011, moreels_etal_2013, moreels_etal_2015, grant_etal_2015, keys_etal_2018}. Therefore, this section also acts as a reminder about the limits of using simplified frameworks to infer certain wave modes and plasma parameters (coronal seismology) from low resolution observational data. 

\subsection{Theoretically expected sausage wave modes}

The dispersion relation of sausage wave modes for the simplified model of a homogeneous tube embedded in a homogeneous background is \citep[e.g.][]{roberts_2019_book}
\begin{equation} \label{eq:dispersion_relation}
    \frac{\kappa_\mathrm{i}}{\rho_\mathrm{i}(k_z^2v_\mathrm{A,i}^2-\omega^2)}\frac{I_0'(\kappa_\mathrm{i}R)}{I_0(\kappa_\mathrm{i}R)}=\frac{\kappa_\mathrm{e}}{\rho_\mathrm{e}(k_z^2v_\mathrm{A,e}^2-\omega^2)}\frac{K_0'(\kappa_\mathrm{e}R)}{K_0(\kappa_\mathrm{e}R)}
\end{equation}
with
\begin{equation} \label{eq:kappa}
    \kappa_\mathrm{i/e}^2=\frac{(k_z^2v_\mathrm{s,i/e}^2-\omega^2)(k_z^2v_\mathrm{A,i/e}^2-\omega^2)}{(v_{s,i/e}^2+v_{A,i/e}^2)(k_z^2v_{t,i/e}^2-\omega^2)},
\end{equation}
where $v_\mathrm{A}$ is the Alfv\'{e}n speed, $v_\mathrm{s}$ is the sound speed, $v_\mathrm{t}=v_\mathrm{A}v_\mathrm{s}/(v_\mathrm{A}^2+v_\mathrm{s}^2)^{1/2}$ is the tube speed, $R$ is the radius of the tube, $k_z$ is the vertical wave number, and $I_0$ and $K_0$ are the modified Bessel functions. The prime in $I_0'$ and $K_0'$ denotes the derivative of the modified Bessel functions with respect to their argument. Subscripts $i$ denote internal values ($r<R$), whereas subscripts $e$ denote external values ($r>R$).

The comparison of the simulation model to the wave modes of the simplified homogeneous cylinder is done locally as a function of $z$. This allows us to take the effects into account that result from the gravitational stratification of the simulation model, however, gravity is not included in the wave equations for the simplified model. Including the gravitational terms in the equations leads to a cutoff frequency for sausage waves \citep{zurbriggen_etal_2020}. The vertical wave number $k_z$ in the theory then corresponds to a parallel wave number $k_\parallel$ for the simulation model. In the following, we denote this wave number as $k$.

The comparison to the simplified model requires the definition of constant internal and external plasma parameters. This can be approximately achieved by averaging the parameters of the initial model ($t=0$ s) within and outside the loop radius $R$, respectively, as a function of $z$. Our model, however, smoothly transitions from loop interior to exterior without an obvious boundary. The determination for the loop radius is therefore afflicted with uncertainties. Field line 2 is rooted at the bottom of the domain at the FWHM of the total magnetic field and was used as the definition for the loop radius in Section \ref{sec:5_energy_trans_into_corona}. Calculating the FWHM of the total magnetic field as a function of $z$, however, results in a far larger radius for larger $z$. Averaging the plasma parameters using either definition for the radius results in similar internal and external values, especially for the bottom 10 Mm we focus on in this study. Therefore, all the following calculations assume field line 2 to define the loop radius.

Technically, determining $v_\mathrm{A,i/e}$, $v_\mathrm{s,i/e}$, and $\rho_\mathrm{i/e}$ by averaging internal and external values results in a model that is not in total pressure balance and thus not in equilibrium. However, when instead setting $\rho_\mathrm{i/e}$ according to the equilibrium condition, the modification of our results is only very minor and can thus be ignored.

\begin{figure}
    \centering
    \includegraphics[width=0.5\textwidth]{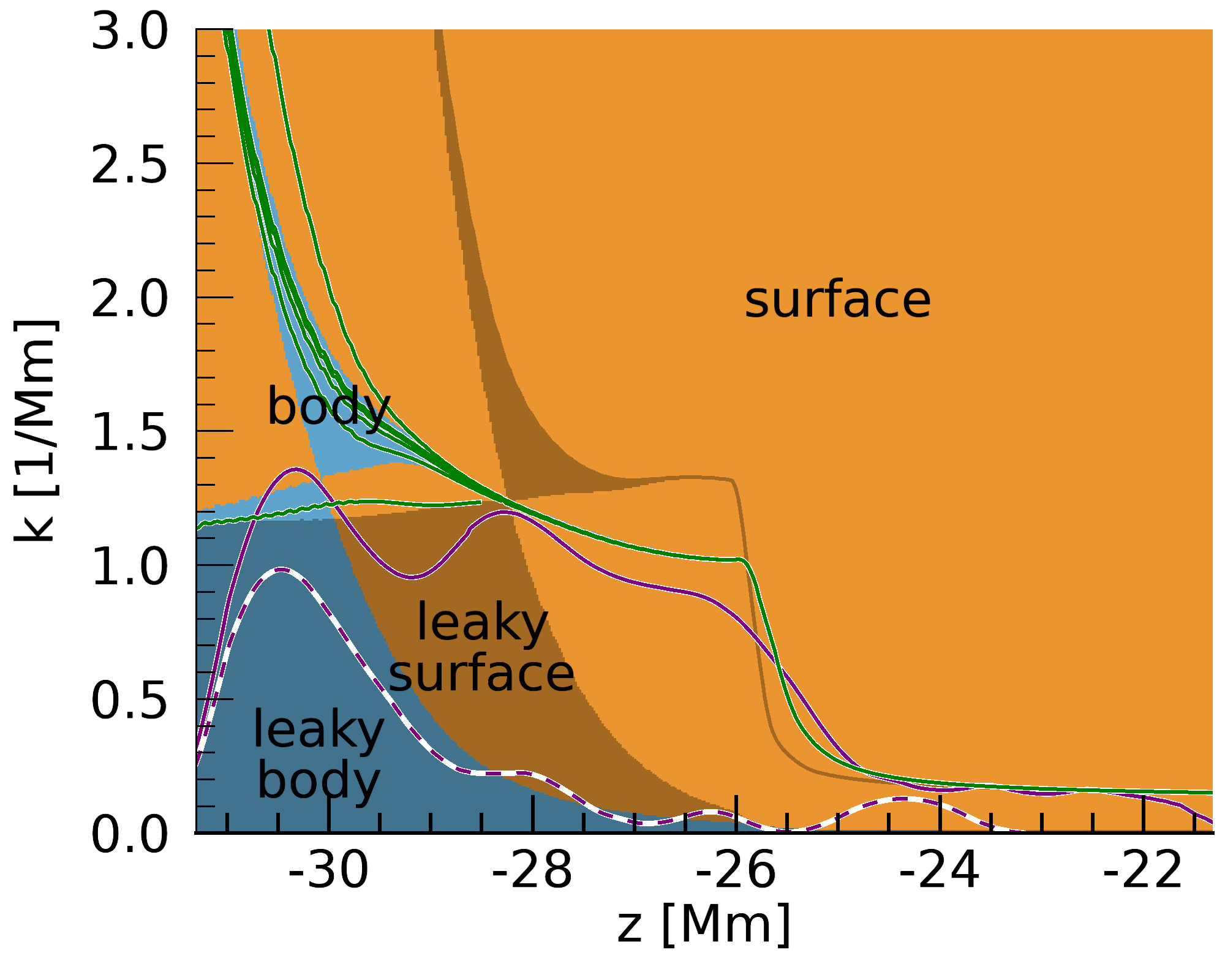}
    \caption{Wave mode regions for sausage waves as a function of vertical/parallel wave number $k$ and $z$ for the simplified model. The overplotted green lines show solutions of the dispersion relation for trapped waves. The purple lines show the phase speeds occurring in the simulations and determined in Figure \ref{fig:5_measured_phase_speed}.}
    \label{fig:5_wave_mode_regions}
\end{figure}

The square of the radial wave number $\kappa$ is given by Equation \ref{eq:kappa}. The nature of $\kappa^2$ determines the wave mode, where for surface waves $\kappa_\mathrm{i}^2>0$, for body waves $\kappa_\mathrm{i}^2<0$ and for trapped (non-evanescent) waves $\kappa_\mathrm{e}^2>0$. For leaky waves, $\kappa_\mathrm{i/e}^2$ is complex, because either $k$ and/or $\omega$ is complex \citep{cally_1986, wilson_1981}. Taking $\omega=2\pi/T$ equal to the driver frequency, we determine the different wave mode regimes based on the sign of $\kappa_\mathrm{i/e}^2$. These wave mode regions are visualized in Figure \ref{fig:5_wave_mode_regions} by patches of different colors. Due to the change in plasma parameters the possible wave modes change significantly with height.

Given the determined internal and external plasma parameters of our model and the driver frequency, we numerically solve the dispersion relation for trapped sausage waves (Equation \ref{eq:dispersion_relation}) to find which wave modes are expected in our model according to the theory. The solutions are plotted in Figure \ref{fig:5_wave_mode_regions} as green lines. Below $z\approx -28$ Mm two solutions for surface waves are found: a fast surface wave for $k\approx 1.2$ Mm$^{-1}$ and a slow surface wave for larger $k$. The fast surface wave solution transitions into a fast body wave solution for heights below $z\approx -30$ Mm. In the slow body wave regime, several solutions are found that correspond to the fundamental mode and radial overtones. For $z>-28$ Mm the body wave solutions are confined between the narrow regime defined by $\omega/v_\mathrm{s,i}$ and $\omega/v_\mathrm{t,i}$, that is enclosed on both sides by a surface mode regime without solutions of the dispersion relation.

A drawback in the representation of wave mode regions of Figure \ref{fig:5_wave_mode_regions} is the assumption of a constant frequency $\omega$ for all displayed heights. However, as was mentioned in Section \ref{sec:5_interaction_of_driver_waves_with_background} of this chapter, there might be other periods present. Plotting the same graph for a frequency corresponding to the period of 300 s, as opposed to the 370 s assumed for the shown figure, only slightly shifts and distorts the regions. Qualitatively, the regions and found solutions of the dispersion relation remain the same.

\begin{figure}
    \centering
    \includegraphics[width=0.4\textwidth]{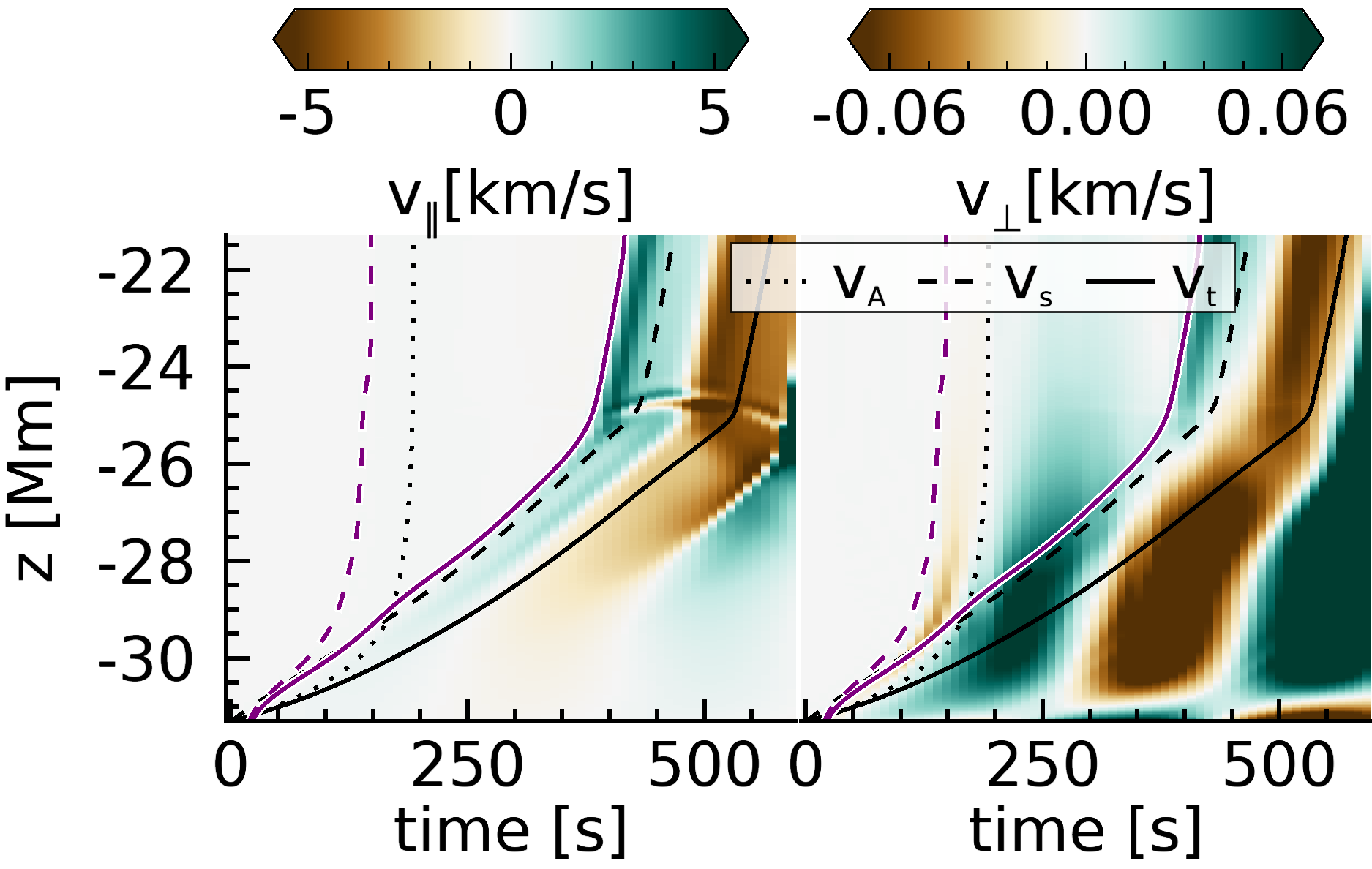}
    \caption{Height-time diagram along field line 1 for the velocity perturbation parallel (left) and normal (right) to the magnetic field. The black lines show the Alfv\'{e}n speed (dotted), the sound speed (dashed), and the tube speed (solid). The purple lines mark two different phase speeds visible in these plots and correspond to the purple lines in Figure \ref{fig:5_wave_mode_regions}. }
    \label{fig:5_measured_phase_speed}
\end{figure}

To measure the wave number $k$ present in our simulations, we need to determine the phase speed. We do so by tracing the wave features of the velocity perturbation in the height-time data along field line 1, which is located close to the loop axis, and calculating the numerical derivative of these lines. Following diverse wave features, we determine two different distance-time lines, which are plotted in Figure \ref{fig:5_measured_phase_speed} as purple lines. The solid line is determined by the onset of the first main wave in the parallel velocity data (left panel), which is not yet subjected to reflections in the cutoff region. Tracing the wave features at later times would not result in the phase speed of the waves, because the waves get altered to standing waves in the cutoff region, which are a superposition of upwards and downwards propagating waves. The dashed line traces the onset of the low amplitude low-$\beta$ fast wave visible in the normal velocity data (right panel). Due to errors in the wave feature tracing and smoothing effects the phase speeds at the bottom of the figure below $z\approx -30.5$ Mm are overestimated. For perfect tracing, the purple lines would start at the lower left corner of the height-time plots.

The two determined phase speeds are plotted with $k=\omega/v_\mathrm{phase}$ in Figure \ref{fig:5_wave_mode_regions} as purple lines with the corresponding linestyles. The oscillations, most prevalent for the solid line, arise due to the slightly wobbly wave tracing in Figure \ref{fig:5_measured_phase_speed} and the subsequent numerical derivative. Ignoring the low values for $k$ at the left side of the plot due to the overestimated phase speed, the solid line is close to the theoretical solution of the fast body and fast surface mode below $z\approx -28$ Mm. Above that height it corresponds well to the theoretical solutions of slow body modes. 

The dashed line in Figure \ref{fig:5_wave_mode_regions} is clearly located in the leaky regime for most of the displayed height range. This line shows the phase speed of a low amplitude wave excited by the initial push of the driver and its leaky nature is no surprise. The leaky nature of that wave can also be seen by the decrease in velocity amplitude in the right panel of Figure \ref{fig:5_measured_phase_speed}, despite the gravitational stratification. Since we have not solved the dispersion relation for leaky waves, we are unable to determine if leaky waves have a theoretical solution close to the dashed line or the lower part of the solid line.

The solid purple line in Figure \ref{fig:5_wave_mode_regions} might hint to the presence of a sausage body mode, that first converts to a sausage surface mode and then again to a sausage body mode. Such a conversion from fast modes to slow modes can theoretically occur if the branches are close in the $k-\omega$ space \citep{cally_2006, khomenko_collados_2006}. However, this line also approximately follows the internal sound speed $v_\mathrm{s,i}$, which would be expected from a plane wave that has not been converted to a sausage wave. Thus, at this point it is impossible to conclude if sausage modes have been excited in our simulations.

A major difficulty in the comparison of the simplified model with the complex simulation model is the difference in the radial (cross sectional) profiles of plasma parameters. While the simplified model, to which the theoretical calculations are applied, is composed of a homogeneous interior and exterior with a discontinuity in between, the simulation model described in Section \ref{subsec:5_model} has continuous radial dependencies without a sudden boundary. For non-uniform tubes without a discontinuity at the boundary there is no clear distinction between surface and body modes and complex $\kappa^2$ can also be found for non-leaky modes due to resonant damping \citep{stenuit_etal_1998,stenuit_etal_1999}. A complex radial wave number $\kappa^2$ can even be found for homogeneous tubes in the presence of electrical resistivity \citep{geeraerts_etal_2020}. 
\citet{soler_etal_2013} formulate a dispersion relation for non-uniform layers of arbitrary thickness between loop interior and exterior and discuss the differences in the eigenfunctions and energy distribution depending on the thickness and the density profile. However, \citet{stenuit_etal_1998} also note that despite not being able to clearly distinguish surface from body modes, there is still a one-to-one correspondence between the modes found for a homogeneous tube and the modes in an inhomogeneous tube for sausage waves. 

Taking this into account, we re-define the wave mode regions plotted in Figure \ref{fig:5_wave_mode_regions}. There, we have assigned the different regions according to the sign of $\kappa_\mathrm{i}^2$ and $\kappa_\mathrm{e}^2$, which were calculated from Equation \ref{eq:kappa} by using the averaged phase speeds inside and outside the loop, respectively. Without averaging, $\kappa^2$ is a function of the $r$-coordinate and may change signs several times, both within and outside the loop. To determine how surface-like or body-like waves in a certain region are expected to behave, we calculate the fraction of area inside the loop with $\kappa^2<0$. A fraction of zero therefore indicates a full surface wave, whereas a fraction of unity indicates a full body wave. Between those values sausage waves have mixed properties. Likewise, we conduct the same calculation for locations outside the loop, which suggests how leaky a sausage wave in a certain regime is.

\begin{figure}
    \centering
    \includegraphics[width=0.5\textwidth]{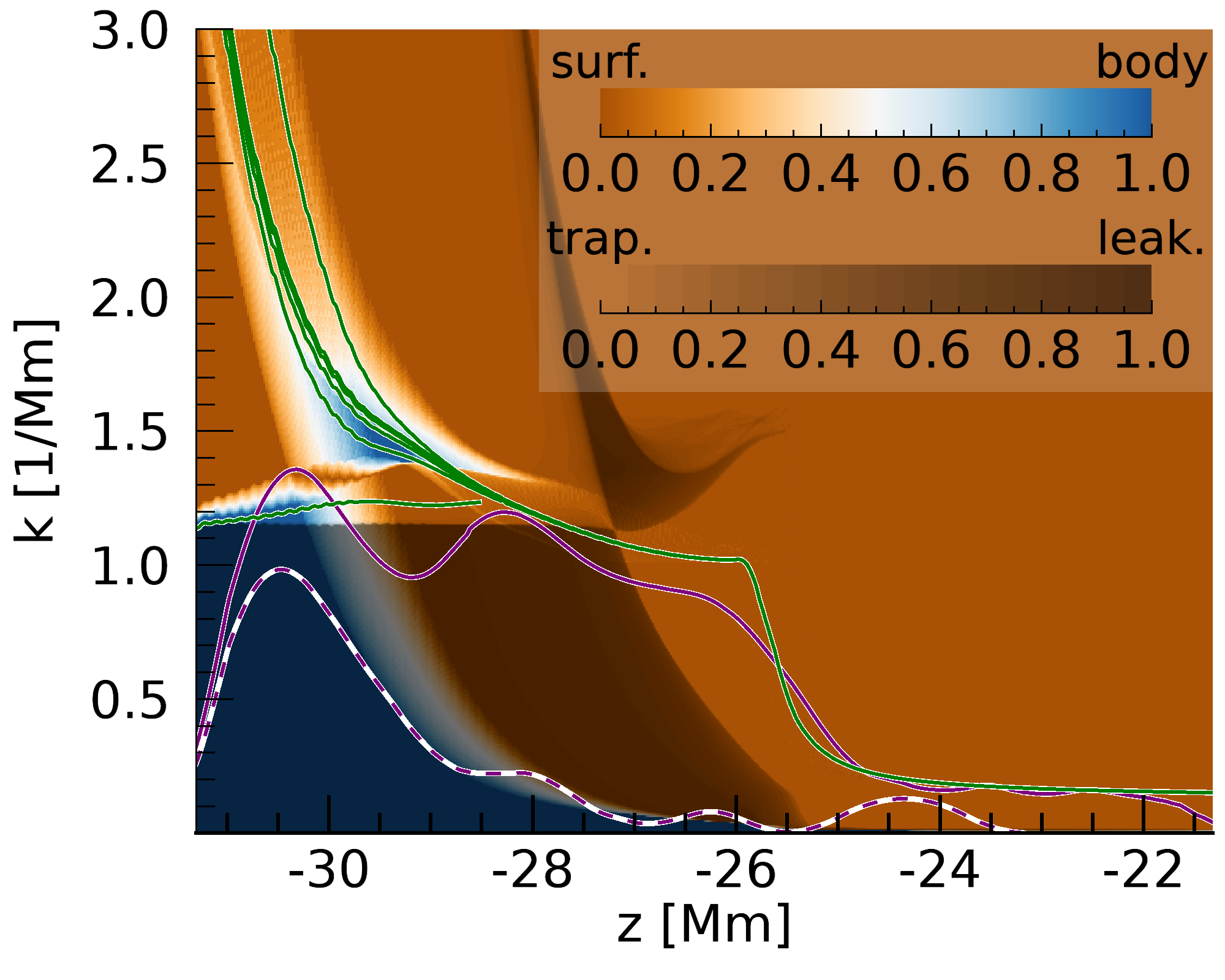}
    \caption{Same as Figure \ref{fig:5_wave_mode_regions}, but the wave mode regions are defined by the fraction of area inside the loop with negative $\kappa^2$ (color) and fraction of area outside the loop with negative $\kappa^2$ (shade). The solutions of the dispersion relation (green lines) still assume a homogeneous tube embedded in a homogeneous background.}
    \label{fig:5_wave_mode_regions_blurry}
\end{figure}

Figure \ref{fig:5_wave_mode_regions_blurry} shows the wave mode regions using this new formalism. The plot shows great similarities with the simpler Figure \ref{fig:5_wave_mode_regions}, with the wave mode regions slightly shifted. As expected, there is a gradual transition between surface and body modes without sharp boundaries. The leaky regime, however, is mostly very clearly defined. The solutions of the dispersion relation (green lines) are the same as in Figure \ref{fig:5_wave_mode_regions} and assume homogeneous $\kappa_\mathrm{i/e}^2$. Due to the slight shift of wave mode regions for considering the radial profile, these lines are slightly off. Still, it is obvious that the slow modes at low heights ($z< -29$ Mm, $k>1.3$ Mm$^{-1}$) are mostly located in a region with mixed surface and body mode properties, whereas the solution for the fast waves ($z<-28$ Mm, $k<1.2$ Mm$^{-1}$) transitions from fully body-like to fully surface-like. The solution for the slow body modes at greater heights ($z>-29$ Mm) are located in a very narrow body mode region, that is enclosed at both sides by a surface mode region. However, due to insufficient resolution this narrow region, as well as the narrow region for leaky surface waves seen for $z>-26.5$ Mm in Figure \ref{fig:5_wave_mode_regions}, is not visible in Figure \ref{fig:5_wave_mode_regions_blurry}.

While Figure \ref{fig:5_wave_mode_regions_blurry} is certainly more advanced than the simpler Figure \ref{fig:5_wave_mode_regions}, the underlying equations still assume the simplified model of a homogeneous tube within a homogeneous background plasma. The comparison between this simplified model and our simulation model is motivated by the fact that also observational studies use the theory of homogeneous flux tubes to identify wave modes. This is usually done for simplicity. However, there also exists a range of spectral codes that can numerically compute normal modes for radially continuously varying equilibria, such as \textsc{Erato} \citep{gruber_etal_1981}, \textsc{Leda} \citep{kerner_etal_1985}, \textsc{Castor} \citep{kerner_etal_1998}, and \textsc{Phoenix} \citep{blokland_etal_2007b}. Recently, a new code, \textsc{Legolas}, has been developed \citep{claes_etal_2020}, which allows the user to solve the full spectra and eigenfunction calculations for any 1D equilibrium, including non-ideal effects. Thus, future work could significantly improve the mode analysis by applying \textsc{Legolas} locally along the $z$-axis on the plasma parameters of our simulation model, which would allow us to take the radial profile into account.

%
%
%

\subsection{Flux ratio inside the loop}

\begin{figure}
\centering
\includegraphics[width=0.4\textwidth]{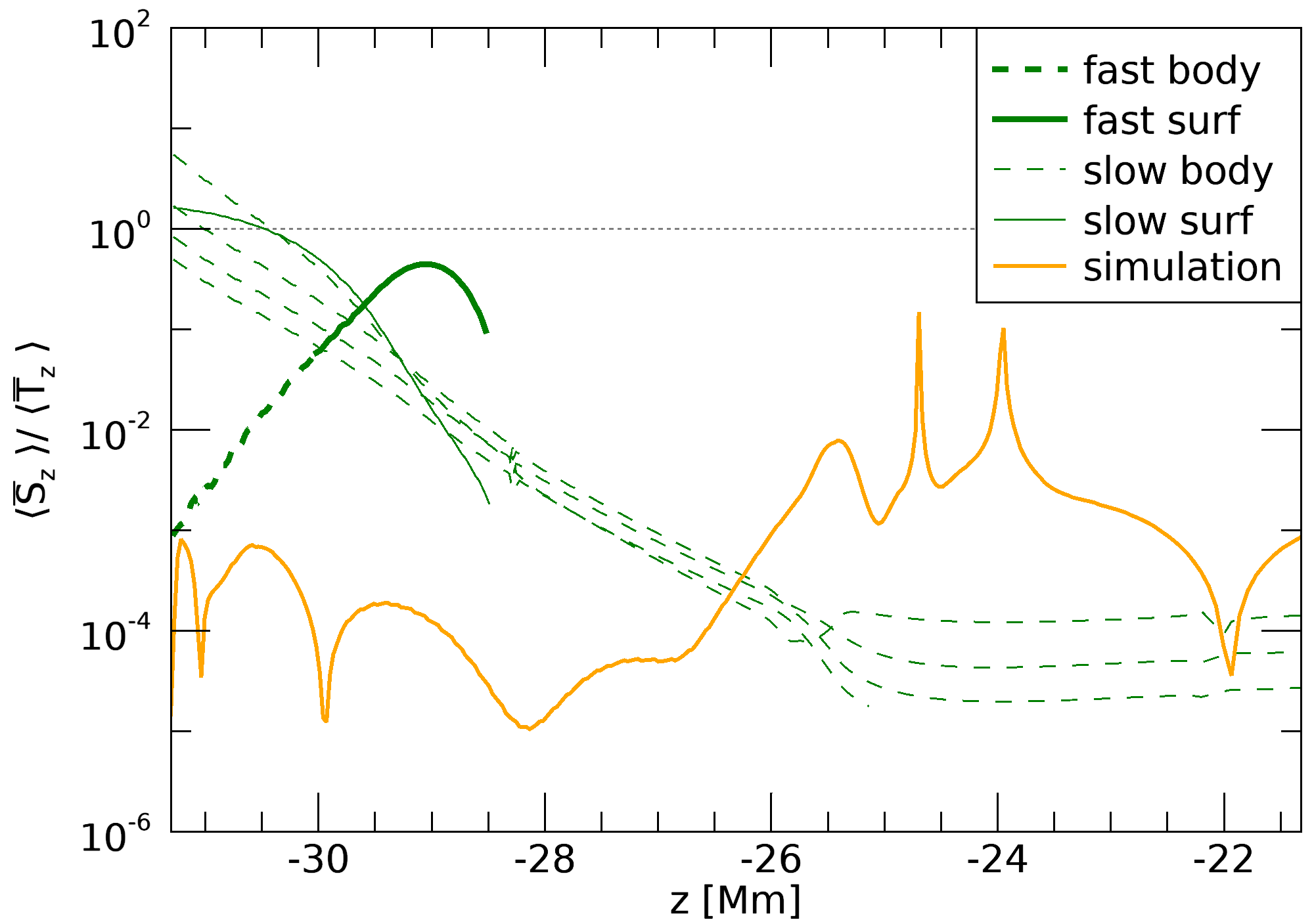}
\caption{Ratio of integrated wave fluxes as a function of height. The green lines show the ratio $\langle \bar{S_z}\rangle / \langle \bar{T_z}\rangle$ inside the loop according to the framework of \citet{moreels_etal_2015} for the theoretical wave modes determined by the solution of the dispersion relation. Thick (thin) lines show fast (slow) modes, while dashed (solid) lines depict body (surface) modes. The solid orange line shows the numerically determined ratio $|\langle \bar{S_\parallel}\rangle| / |\langle \bar{F}_\mathrm{HD,\parallel}\rangle|$ of the simulation results inside the loop.}
\label{fig:5_ratio_of_integrated_fluxes}
\end{figure}

\citet{moreels_etal_2015} have developed a framework to calculate the energy and fluxes of sausage modes within loops using observational parameters. They present analytical formulas to calculate the Poynting flux and thermal flux averaged over one wave period. Then, they integrate the fluxes, which only have a vertical component for trapped waves, over the entire loop cross section. We apply this analytical framework to determine the ratio of the integrated time-averaged Poynting flux and the thermal flux  $\langle \bar{S_z}\rangle / \langle \bar{T_z}\rangle$ on the theoretical modes calculated by solving the dispersion relation. The angled brackets depict the time average, whereas the bar denotes the cross sectional integration. This ratio is displayed in Figure \ref{fig:5_ratio_of_integrated_fluxes} for the different wave modes as green lines. 

According to linear MHD theory, the total flux consists of the sum of the Poynting flux and the thermal flux. For the non-linearized equations, however, the total flux (Equation \ref{eq:5_wave_flux}) has additional terms and consists of the Poynting flux $\Vec{S}$ (first term of Equation \ref{eq:5_wave_flux}) and the HD part $\Vec{F}_\mathrm{HD}$ (right terms). We calculate the wave flux parallel to the magnetic field according to Equation \ref{eq:5_wave_flux} using the perturbed quantities from the simulation data and integrate it over the loop cross section, after averaging it in time over the last period (from $t=3T$ to $t=4T$) of the simulation time series. The ratio $|\langle \bar{S_\parallel}\rangle| / |\langle \bar{F}_\mathrm{HD,\parallel}\rangle|$ is displayed in Figure \ref{fig:5_ratio_of_integrated_fluxes} as orange line. 

We stress that the flux ratios obtained from the solution of the dispersion relation (green lines) only correspond to trapped sausage modes, whereas the flux ratios from the simulation data (orange line) were obtained at a time where the waves are leaky between $z\approx -31$ Mm and $z\approx -29$ Mm. Thus, in this height range a comparison is difficult. Above $z\approx -29$ Mm the theory and simulation data still do not correspond well, but both show very low values for the Poynting flux compared to the thermal or HD flux. We deem a slow body mode in the simulations for those heights as still plausible. At very low heights below $z\approx -31$ Mm the numerical flux ratio is close to the theoretical flux ratio for a fast body wave.

\subsection{Ratio of plasma velocities}

Using the framework of \citet{moreels_vandoorsselaere_2013} it is possible to determine the nature of sausage and kink waves in simulations and observations by the comparison of the ratio of absolute Lagrangian plasma displacements parallel and normal to the magnetic field. This was done for the simulations of \citet{riedl_etal_2019}. For this study, we repeat that investigation, but we substitute the ratio of plasma displacements by the ratio of perturbed velocities, which results in exactly the same expressions. The necessary equations for the velocity eigenfunctions can be found in \citet{roberts_2019_book}.

\begin{figure}
\centering
\includegraphics[width=0.5\textwidth]{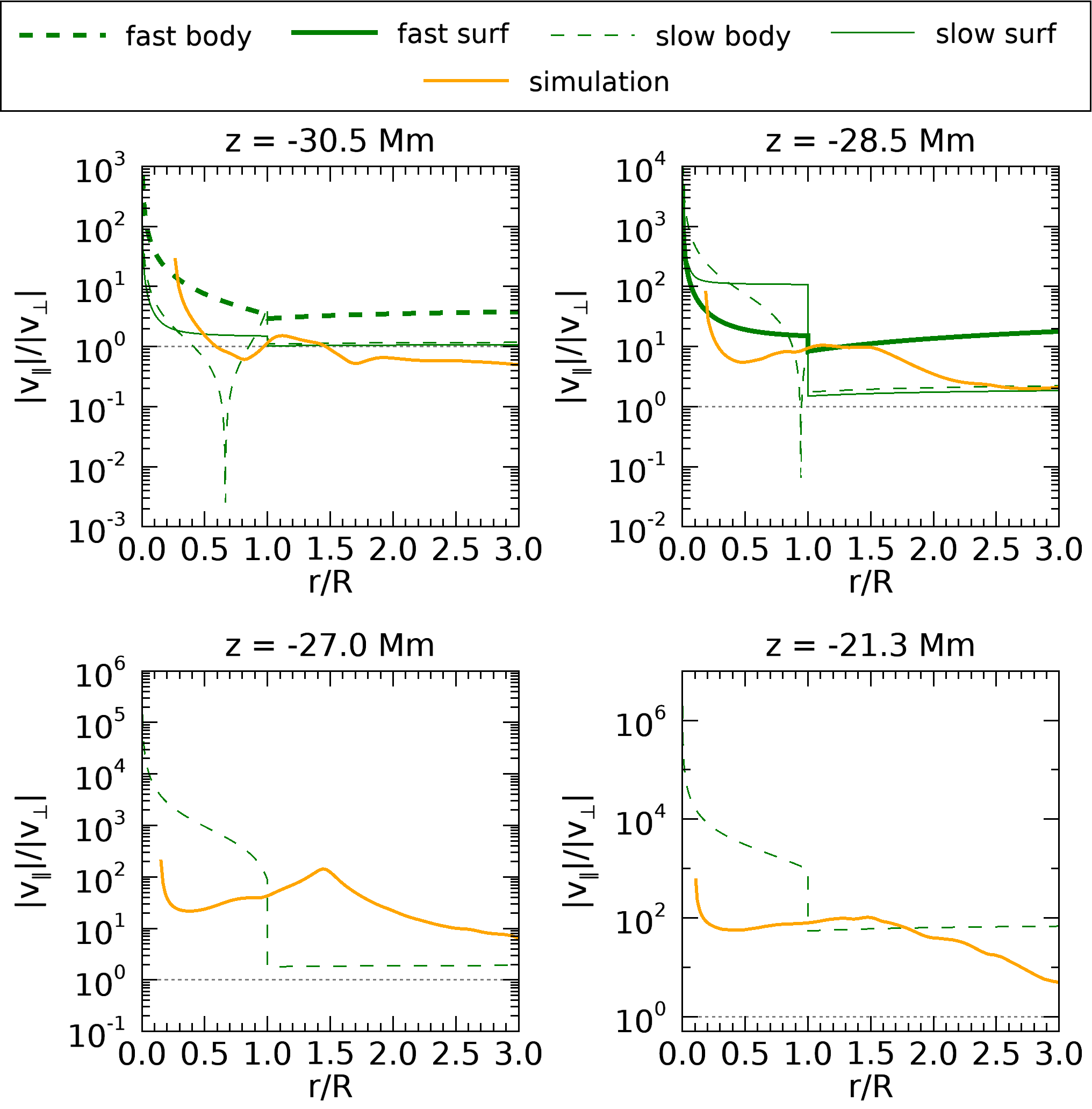}
\caption{Ratio of absolute parallel to normal velocity perturbations for four different heights as a function of normalized radius. Green lines correspond to theoretical values using the solution of the dispersion relation. The linestyles are the same as in Figure \ref{fig:5_ratio_of_integrated_fluxes}. For the slow body mode only the fundamental mode is plotted. The orange lines show the ratio of mean absolute parallel to normal velocity perturbations for the simulation data for the last period ($t=3T$ to $t=4T$).}
\label{fig:5_velocity_ratios}
\end{figure}

The ratio of theoretical velocity perturbations are plotted as green lines in Figure \ref{fig:5_velocity_ratios} for four different heights. For the slow body mode only the fundamental mode is plotted. The velocity ratios of the radial overtones for this mode show several strong spikes within the loop ($r/R<1$) due to nodes. This behavior is not seen in the simulations. For the simulation data (orange lines in Figure \ref{fig:5_velocity_ratios}), we use the ratio of the mean absolute velocity perturbation $\langle|v_\parallel|\rangle/\langle|v_\perp|\rangle$. 

The top panels of Figure \ref{fig:5_velocity_ratios} are located at heights where the waves in the simulation may already (left) or still (right) be of leaky nature, which makes a comparison to the theoretical velocity ratios for trapped waves difficult. These panels are included for comparison. 
The bottom panels display the data at heights well in the propagating regime where theoretically only slow body modes are present. The bottom left panel is located below the TR, while the bottom right panel is located above the TR. There are significant differences between the theory and the simulations for these panels, which decreases the probability of a slow body mode at those heights. Of course, the comparison between the theoretical curves for the simplified model and the ones obtained from the simulation data have to be made with caution, because the eigenfunctions change if the non-uniformity of the loop interior and the background are considered \citep{soler_etal_2013}. In addition, there is no discontinuity in the simulation model between internal and external values, so the absence of sharp jumps of the orange curve around $r/R=1$ is not surprising, even if sausage waves have been excited.

\subsection{Further considerations for wave mode determination}

As \citet{keys_etal_2018} accurately summarized, surface and body modes can be distinguished by the behavior of their eigenfunctions inside the tube. The vertical amplitude for surface waves, for example, decays exponentially for increasing distances from the boundary, while this is not the case inside the tube for body waves. Body waves also can have several nodes inside the tube. However, we remind the reader that, as mentioned above, for the case of an inhomogeneous tube without a clear boundary the distinction between surface and body waves is not clear \citep{stenuit_etal_1998, stenuit_etal_1999}.

\begin{figure}
\centering
\includegraphics[width=0.8\textwidth]{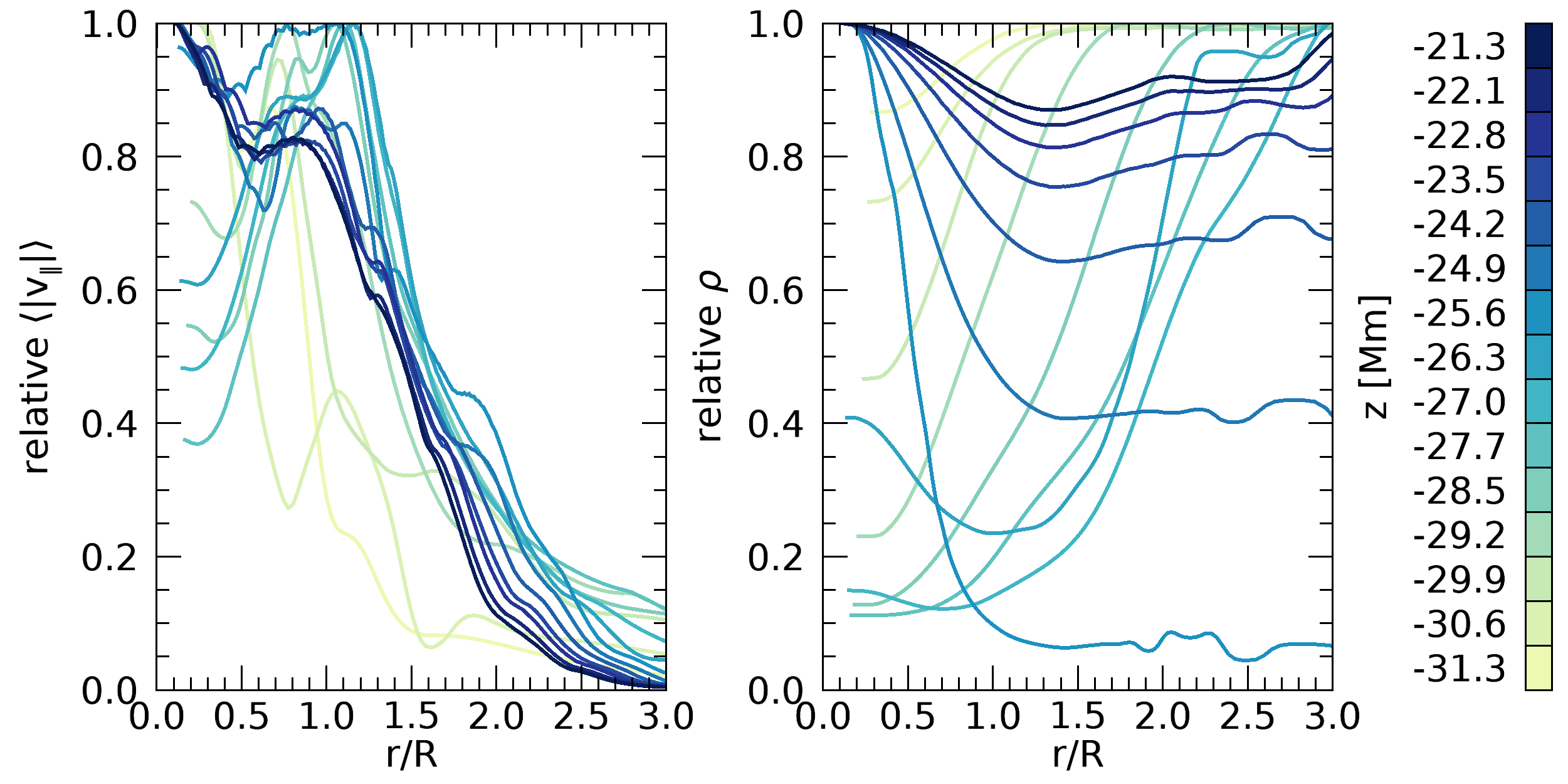}\vspace{0.3cm}
\includegraphics[width=0.8\textwidth]{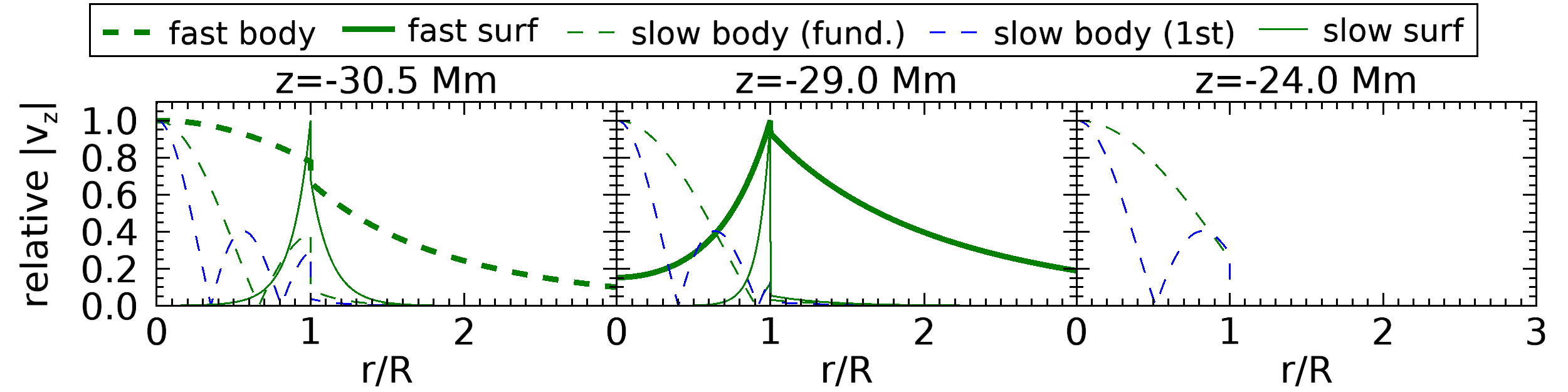}
\caption{\textit{Top left:} Absolute parallel velocity perturbation averaged from $t=3T$ to $t=4T$ as a function of normalized radial coordinate for different heights for the simulation with $t_0=0$ s. The velocity is normalized to the maximum value at every height in the displayed range. \textit{Top right:} Cross section of the initial density of the simulated model (at $t=0$ s) for different heights. The normalization is done as in the left plot. \textit{Bottom:} Absolute amplitude of the vertical velocity according to the eigenfunctions for three different heights. For the slow body mode, the fundamental mode and the first radial overtone are plotted.}
\label{fig:5_vp_mean+rho_with_height}
\end{figure}

The top left panel of Figure \ref{fig:5_vp_mean+rho_with_height} shows the vertical velocity amplitude for the simulation as a function of radius for different heights. While close to the bottom of the domain and above the TR the maximum amplitude is clearly located at the loop axis, there is a striking increase in amplitude around the location of the loop boundary ($r/R=1$) between the upper end of the main cutoff region and the TR. While the first could be attributed to body waves, the second might be a sign of surface waves. Another possibility for this shape of the parallel velocity amplitudes apart from sausage waves may be wave refraction owing to the complicated and time-dependent sound speed and Alfv\'{e}n speed profiles. However, while some influence from wave refraction cannot be ruled out, the fact that the peaks we attribute to sausage surface waves all occur at $r/R=1$ clearly shows that the waves are influenced by the loop boundary. The sound and Alfv\'{e}n speed profiles do not show any special shapes around field line 2, which was used to define the loop boundary. Finally, we also have to account for the fact that a cutoff region for field-aligned wave propagation is only present for the inner part of the loop, which could explain why the velocity perturbations are higher outside this region which coincides with field line 2 (see Figures \ref{fig:5_cutoff_region} and \ref{fig:5_parallel_HD_flux_time_series}). However, this would neither explain why the waves still show a surface-like nature well above the main cutoff region, nor why the maximum parallel velocity perturbation is located again at the loop axis for greater heights. Thus, the changing behavior of the relative parallel velocity perturbation amplitude as a function of height is difficult to explain for plane waves, which hints to the presence of sausage modes. An exception has to be noted for the parallel velocity profile at low heights, which could very well be explained by plane waves due to the Gaussian shape of the driver.


The bottom panels of Figure \ref{fig:5_vp_mean+rho_with_height} show for comparison the absolute vertical velocity amplitude expected from the eigenfunctions for the different sausage wave modes. We note that the heights for which these eigenfunctions are plotted do not exactly correspond to where we find sausage-like or body-like behavior in the simulations. This is because due to the differences in the simulation model and the homogeneous simplified model the wave mode regions in Figures \ref{fig:5_wave_mode_regions} and \ref{fig:5_wave_mode_regions_blurry} are expected to be shifted and deformed. Nonetheless, the three bottom panels of \ref{fig:5_vp_mean+rho_with_height} show clear similarities to the amplitudes appearing in the simulations, namely a (fast or slow fundamental) body mode close to the bottom of the domain, a (fast or slow) surface mode above, followed by a slow (fundamental or first harmonic) body mode. Eigenfunctions of leaky waves, which are expected to occur in the simulation for heights between $z\approx  -31$ Mm and $z\approx -29$ Mm, are not plotted.

The question arises if and where surface-like waves are even plausible for our simulation model with a smooth loop boundary. For kink waves, surface waves disappear for media with broad transition layers between internal and external plasma parameters \citep{van_doorsselaere+poedts_2007}. The top right panel of Figure \ref{fig:5_vp_mean+rho_with_height} shows the radial density profile for our model for different heights. Surface waves would be expected where the boundary is sufficiently sharp. The density contrast between loop interior and exterior, and therefore also the steepness of the boundary, is greatest for heights where surface-like behavior is found in the top left panel. This supports the possibility of surface-like sausage waves for those heights. As a side note we would like to mention that the density profile reverses in the TR, which can also be seen in the right panel of Figure \ref{fig:5_full_domain}.

To summarize, mode identification is very difficult for complex loop models and - considering the unclear distinction between surface and body modes already encountered in theory for much simpler inhomogeneous tubes \citep{stenuit_etal_1998, stenuit_etal_1999} - even impossible. Moreover, there is a possibility that different surface-like or body-like modes are present at the same height at the same time and superimposed, which would be very difficult to distinguish from higher harmonic body modes. Simultaneous sausage and kink modes in the chromosphere were already observed by \citet{morton_etal_2012}, which makes the simultaneous presence of several kinds of sausage modes likely. However, we have made an effort to identify the most prevalent modes appearing in our simulations. Depending on which wave property is investigated (phase speed, fluxes, or eigenfunctions) the waves in our simulations agree more or less with the localized linear theory for sausage waves in a homogeneous cylinder. Therefore, it remains inconclusive if sausage waves are actually excited in our simulations or if the waves retain their plane character. If sausage waves are indeed (partly) present, our results indicate a (possibly leaky) body-like mode at low heights, that converts to a (possibly leaky) surface-like mode above the main cutoff region, and then into a (trapped) slow body-like mode above the TR. Following the solution of the dispersion relation displayed in Figures \ref{fig:5_wave_mode_regions} and \ref{fig:5_wave_mode_regions_blurry}, the first body-like mode and the surface-like mode are likely fast or leaky waves. The difficulties to identify modes in the present simulations highlights the challenges observers have to face, who do not have the same luxury of both high resolution and knowledge about the involved plasma parameters.


\section{Conclusions} \label{sec:5_conclusions}

We conducted numerical 3D MHD simulations of a whole stratified loop spanning from photosphere to photosphere using the \textsc{Pluto} code \citep{mignone_etal_2007,mignone_etal_2012,mignone_etal_2018}. As initial condition we used the numerically relaxed model of \citet{reale_etal_2016}. The loop significantly expands in the chromosphere and does not only show vertical gravitational stratification, but also horizontal structuring. We applied a localized vertical acoustic-gravity driver inside the loop at one footpoint and investigated the resulting wave behavior. The driver has a period of $T=370$ s, which is close to the typical peak periods excited by photospheric \textit{p}-modes. 

Following the nature of the driver, most of the wave energy flux remains acoustic and there is only very little conversion to magnetic properties (Poynting flux). This agrees with the findings of \citet{riedl_etal_2019} for the case of vertical flux tubes. Compared to the study of \citet{riedl_etal_2019} the identification of wave modes was even more difficult. It has to be assumed that a clear distinction between modes is not possible and that the different wave mode regions defined by the sign of the squared radial wave number $\kappa^2$ are blurred, as can be seen by comparing Figures \ref{fig:5_wave_mode_regions} and \ref{fig:5_wave_mode_regions_blurry}. The simulation results indicate that the initially plane waves excited by the driver may convert first to a fast or leaky sausage body-like mode, that transitions to a fast or leaky sausage surface-like mode at the top of the main cutoff region, and converts to a slow body-like mode above the TR. The transition between modes is expected to be smooth, because of the smooth radial dependence of the plasma parameters in the simulation model. According to the solution of the dispersion relation for sausage waves in a homogeneous cylinder such a conversion between body-like and surface-like waves is possible due to the change of plasma parameters with height. The upper end of the fast surface wave branch is in very close proximity to the branch of slow body modes, which makes a conversion from one to the other plausible. To the best of our knowledge, such a conversion between sausage modes has not yet been observed on the Sun. An attempt was very recently made by \citet{grant_etal_2021_in_prep} for solar pores. However, the observed pores that showed body modes in the photosphere had a weak magnetic field and were therefore not reliably detectable in chromospheric heights. This poses a serious limitation in the observability of sausage surface to body mode transformation in solar pores.

\citet{riedl_etal_2019} also found a fast sausage surface wave and a slow sausage body wave for vertical flux tubes. The excitation of kink waves and conversion of wave flux from acoustic to magnetic properties for inclined flux tubes was not studied in this paper, because unlike in the study of \citet{riedl_etal_2019} the loop could not be rotated. However, this issue can be effectively resolved by applying a driver with an oblique polarization. Furthermore, kink waves may also be excited if the symmetry of the system is broken by shifting the position of the driver, so that the peak of the Gaussian is not located at the loop axis. Possibly, applying a driver with higher frequency would facilitate the identification of wave modes due to the shorter wavelength in comparison to the scale height. In addition, improvements could be made by also comparing the waves in the simulations to leaky sausage waves and by accounting for the smooth radial profiles of the plasma parameters.

The presence of acoustic cutoff regions significantly alters the wave behavior. Inside the main cutoff region predicted by a localized version of the isothermal acoustic cutoff frequency proposed by \citet{lamb_1909}, standing waves are found. This is as expected and validates the applicability of this simple formula. A significant part of wave energy is still tunneling through and reaches greater heights. Considering only the waves itself, that are excited by the driver, the waves are propagating above the main cutoff region with approximately the sound speed. Close to the TR, we observed the formation of standing waves, which agrees with the presence of a second cutoff region predicted by the formulas of \citet{deubner_gough_1984}, \citet{schmitz_fleck_1998}, and \citet{roberts_2006}. Taking the full energy flux into account, which includes all background motions present due to the imperfectly relaxed model, we find standing waves in the whole region between the main cutoff region and the TR. These standing waves appear due to the interaction of the driven waves with the background motions and lead to an oscillation of the TR height. 

The cutoff regions and the TR height are highly dynamic, which is in accordance to what is expected and known in reality. The dynamic behavior arises due to an interplay of background motions and driven waves. Different simulations, that only differ in the time when the driver is switched on, clearly show the influence of the driver. While the TR also moves in simulations without a driver, its movement is stronger and has dips and peaks at different times than for the driven simulations. The periodicity of the TR movement is not constant at first, but approaches values similar to the driver frequency for longer run times.

Wave energy flux mainly leaks into the corona at times when the TR is lowest. Only approximately 2\% of the wave energy excited by the driver is transmitted into the corona, or approximately 4\% of the wave energy from just below the main cutoff region. The transmitted flux is mainly of acoustic nature. The cutoff regions play a significant role in the flux damping, however, the geometric damping effects described by \citet{riedl_etal_2021} are also deemed important. The loop significantly expands, which may lead to considerable damping due to geometric spreading. In addition, we confirmed that there is also lateral wave leakage in a 3D system containing a realistic loop. However, we have not yet investigated if the lateral wave leakage corresponds to the so-called leaky waves from the tube wave theory.

\begin{acknowledgments}
JMR and TVD have received funding from the European Research Council (ERC) under the European Union's Horizon 2020 research and innovation programme (grant agreement No. 724326).
FR and AP acknowledge support from Italian Ministero dell'Universit\`{a} e della Ricerca.
\end{acknowledgments}

\bibliography{Paper}{}

\begin{thebibliography}{}
\expandafter\ifx\csname natexlab\endcsname\relax\def\natexlab#1{#1}\fi
\providecommand{\url}[1]{\href{#1}{#1}}
\providecommand{\dodoi}[1]{doi:~\href{http://doi.org/#1}{\nolinkurl{#1}}}
\providecommand{\doeprint}[1]{\href{http://ascl.net/#1}{\nolinkurl{http://ascl.net/#1}}}
\providecommand{\doarXiv}[1]{\href{https://arxiv.org/abs/#1}{\nolinkurl{https://arxiv.org/abs/#1}}}

\bibitem[{{Alexiades} {et~al.}(1996){Alexiades}, {Amiez}, \&
  {Gremaud}}]{alexiades_etal_1996}
{Alexiades}, V., {Amiez}, G., \& {Gremaud}, P. 1996, Communications in
  Numerical Methods in Engineering, 12, 31,
  \dodoi{https://doi.org/10.1002/(SICI)1099-0887(199601)12:1<31::AID-CNM950>3.0.CO;2-5}

\bibitem[{{Anfinogentov} {et~al.}(2015){Anfinogentov}, {Nakariakov}, \&
  {Nistic{\`o}}}]{anfinogentov_etal_2015}
{Anfinogentov}, S.~A., {Nakariakov}, V.~M., \& {Nistic{\`o}}, G. 2015, \aap,
  583, A136, \dodoi{10.1051/0004-6361/201526195}

\bibitem[{{Beck} {et~al.}(2009){Beck}, {Khomenko}, {Rezaei}, \&
  {Collados}}]{beck_etal_2009}
{Beck}, C., {Khomenko}, E., {Rezaei}, R., \& {Collados}, M. 2009, \aap, 507,
  453, \dodoi{10.1051/0004-6361/200911851}

\bibitem[{{Bel} \& {Leroy}(1977)}]{bel_leroy_1977}
{Bel}, N., \& {Leroy}, B. 1977, \aap, 55, 239

\bibitem[{{Beli{\"e}n} {et~al.}(1999){Beli{\"e}n}, {Martens}, \&
  {Keppens}}]{belien_etal_1999}
{Beli{\"e}n}, A.~J.~C., {Martens}, P.~C.~H., \& {Keppens}, R. 1999, \apj, 526,
  478, \dodoi{10.1086/307980}

\bibitem[{{Blokland} {et~al.}(2007){Blokland}, {van der Holst}, {Keppens}, \&
  {Goedbloed}}]{blokland_etal_2007b}
{Blokland}, J.~W.~S., {van der Holst}, B., {Keppens}, R., \& {Goedbloed}, J.~P.
  2007, Journal of Computational Physics, 226, 509,
  \dodoi{10.1016/j.jcp.2007.04.018}

\bibitem[{{Bogdan} {et~al.}(1996){Bogdan}, {Hindman}, {Cally}, \&
  {Charbonneau}}]{bogdan_etal_1996}
{Bogdan}, T.~J., {Hindman}, B.~W., {Cally}, P.~S., \& {Charbonneau}, P. 1996,
  \apj, 465, 406, \dodoi{10.1086/177429}

\bibitem[{{Bogdan} {et~al.}(2003){Bogdan}, {Carlsson}, {Hansteen}, {McMurry},
  {Rosenthal}, {Johnson}, {Petty-Powell}, {Zita}, {Stein}, {McIntosh}, \&
  {Nordlund}}]{bogdan_etal_2003}
{Bogdan}, T.~J., {Carlsson}, M., {Hansteen}, V.~H., {et~al.} 2003, \apj, 599,
  626, \dodoi{10.1086/378512}

\bibitem[{{Botha} {et~al.}(2011){Botha}, {Arber}, {Nakariakov}, \&
  {Zhugzhda}}]{botha_etal_2011}
{Botha}, G.~J.~J., {Arber}, T.~D., {Nakariakov}, V.~M., \& {Zhugzhda}, Y.~D.
  2011, \apj, 728, 84, \dodoi{10.1088/0004-637X/728/2/84}

\bibitem[{{Cally}(1986)}]{cally_1986}
{Cally}, P.~S. 1986, \solphys, 103, 277, \dodoi{10.1007/BF00147830}

\bibitem[{{Cally}(2006)}]{cally_2006}
---. 2006, Philosophical Transactions of the Royal Society of London Series A,
  364, 333, \dodoi{10.1098/rsta.2005.1702}

\bibitem[{{Cally}(2017)}]{cally_2017}
---. 2017, \mnras, 466, 413, \dodoi{10.1093/mnras/stw3215}

\bibitem[{{Cally} \& {Goossens}(2008)}]{cally+goossens_2008}
{Cally}, P.~S., \& {Goossens}, M. 2008, \solphys, 251, 251,
  \dodoi{10.1007/s11207-007-9086-3}

\bibitem[{{Cally} \& {Hansen}(2011)}]{cally+hansen_2011}
{Cally}, P.~S., \& {Hansen}, S.~C. 2011, \apj, 738, 119,
  \dodoi{10.1088/0004-637X/738/2/119}

\bibitem[{{Cally} \& {Khomenko}(2019)}]{cally_khomenko_2019_part_I}
{Cally}, P.~S., \& {Khomenko}, E. 2019, \apj, 885, 58,
  \dodoi{10.3847/1538-4357/ab3bce}

\bibitem[{{Centeno} {et~al.}(2006){Centeno}, {Collados}, \& {Trujillo
  Bueno}}]{centeno_etal_2006}
{Centeno}, R., {Collados}, M., \& {Trujillo Bueno}, J. 2006, \apj, 640, 1153,
  \dodoi{10.1086/500185}

\bibitem[{{Claes} {et~al.}(2020){Claes}, {De Jonghe}, \&
  {Keppens}}]{claes_etal_2020}
{Claes}, N., {De Jonghe}, J., \& {Keppens}, R. 2020, \apjs, 251, 25,
  \dodoi{10.3847/1538-4365/abc5c4}

\bibitem[{{De Moortel} {et~al.}(2002){De Moortel}, {Ireland}, {Hood}, \&
  {Walsh}}]{de_moortel_etal_2002}
{De Moortel}, I., {Ireland}, J., {Hood}, A.~W., \& {Walsh}, R.~W. 2002, \aap,
  387, L13, \dodoi{10.1051/0004-6361:20020436}

\bibitem[{{De Pontieu} {et~al.}(2005){De Pontieu}, {Erd{\'{e}}lyi}, \& {De
  Moortel}}]{de_pontieu_etal_2005}
{De Pontieu}, B., {Erd{\'{e}}lyi}, R., \& {De Moortel}, I. 2005, The
  Astrophysical Journal, 624, L61, \dodoi{10.1086/430345}

\bibitem[{{De Pontieu} {et~al.}(2004){De Pontieu}, {Erd{\'e}lyi}, \&
  {James}}]{de_pontieu_etal_2004}
{De Pontieu}, B., {Erd{\'e}lyi}, R., \& {James}, S.~P. 2004, \nat, 430, 536,
  \dodoi{10.1038/nature02749}

\bibitem[{{Deubner} \& {Gough}(1984)}]{deubner_gough_1984}
{Deubner}, F.-L., \& {Gough}, D. 1984, \araa, 22, 593,
  \dodoi{10.1146/annurev.aa.22.090184.003113}

\bibitem[{{Erd{\'e}lyi} {et~al.}(2007){Erd{\'e}lyi}, {Malins}, {T{\'o}th}, \&
  {de Pontieu}}]{erdelyi_etal_2007}
{Erd{\'e}lyi}, R., {Malins}, C., {T{\'o}th}, G., \& {de Pontieu}, B. 2007,
  \aap, 467, 1299, \dodoi{10.1051/0004-6361:20066857}

\bibitem[{{Fedun} {et~al.}(2011){Fedun}, {Shelyag}, \&
  {Erd{\'e}lyi}}]{fedun_etal_2011}
{Fedun}, V., {Shelyag}, S., \& {Erd{\'e}lyi}, R. 2011, \apj, 727, 17,
  \dodoi{10.1088/0004-637X/727/1/17}

\bibitem[{{Felipe}(2019)}]{felipe_2019}
{Felipe}, T. 2019, \aap, 627, A169, \dodoi{10.1051/0004-6361/201935784}

\bibitem[{{Felipe} \& {Sangeetha}(2020)}]{felipe+sangeetha_2020}
{Felipe}, T., \& {Sangeetha}, C.~R. 2020, \aap, 640, A4,
  \dodoi{10.1051/0004-6361/202038387}

\bibitem[{{Freij} {et~al.}(2016){Freij}, {Dorotovi{\v{c}}}, {Morton},
  {Ruderman}, {Karlovsk{\'y}}, \& {Erd{\'e}lyi}}]{frij_etal_2016}
{Freij}, N., {Dorotovi{\v{c}}}, I., {Morton}, R.~J., {et~al.} 2016, \apj, 817,
  44, \dodoi{10.3847/0004-637X/817/1/44}

\bibitem[{{Gascoyne} {et~al.}(2014){Gascoyne}, {Jain}, \&
  {Hindman}}]{gascoyne_etal_2014}
{Gascoyne}, A., {Jain}, R., \& {Hindman}, B.~W. 2014, \apj, 789, 109,
  \dodoi{10.1088/0004-637X/789/2/109}

\bibitem[{{Geeraerts} {et~al.}(2020){Geeraerts}, {Van Doorsselaere}, {Chen}, \&
  {Li}}]{geeraerts_etal_2020}
{Geeraerts}, M., {Van Doorsselaere}, T., {Chen}, S.-X., \& {Li}, B. 2020, \apj,
  897, 120, \dodoi{10.3847/1538-4357/ab9b28}

\bibitem[{{Gilchrist-Millar} {et~al.}(2021){Gilchrist-Millar}, {Jess}, {Grant},
  {Keys}, {Beck}, {Jafarzadeh}, {Riedl}, {Van Doorsselaere}, \& {Ruiz
  Cobo}}]{gilchrist_etal_2021}
{Gilchrist-Millar}, C.~A., {Jess}, D.~B., {Grant}, S. D.~T., {et~al.} 2021,
  Philosophical Transactions of the Royal Society of London Series A, 379,
  20200172, \dodoi{10.1098/rsta.2020.0172}

\bibitem[{{Grant} {et~al.}(2021){Grant}, {Jess}, \&
  {Jafarzadeh}}]{grant_etal_2021_in_prep}
{Grant}, S.~D.~T., {Jess}, D.~B., \& {Jafarzadeh}, S. 2021

\bibitem[{{Grant} {et~al.}(2015){Grant}, {Jess}, {Moreels}, {Morton},
  {Christian}, {Giagkiozis}, {Verth}, {Fedun}, {Keys}, {Van Doorsselaere}, \&
  {Erd{\'e}lyi}}]{grant_etal_2015}
{Grant}, S.~D.~T., {Jess}, D.~B., {Moreels}, M.~G., {et~al.} 2015, \apj, 806,
  132, \dodoi{10.1088/0004-637X/806/1/132}

\bibitem[{{Gruber} {et~al.}(1981){Gruber}, {Troyon}, {Berger}, {Bernard},
  {Rousset}, {Schreiber}, {Kerner}, {Schneider}, \&
  {Roberts}}]{gruber_etal_1981}
{Gruber}, R., {Troyon}, F., {Berger}, D., {et~al.} 1981, Computer Physics
  Communications, 21, 323, \dodoi{10.1016/0010-4655(81)90013-8}

\bibitem[{{Guarrasi} {et~al.}(2014){Guarrasi}, {Reale}, {Orlando}, {Mignone},
  \& {Klimchuk}}]{guarrasi_etal_2014}
{Guarrasi}, M., {Reale}, F., {Orlando}, S., {Mignone}, A., \& {Klimchuk}, J.~A.
  2014, \aap, 564, A48, \dodoi{10.1051/0004-6361/201322848}

\bibitem[{{Heggland} {et~al.}(2011){Heggland}, {Hansteen}, {De Pontieu}, \&
  {Carlsson}}]{heggland_etal_2011}
{Heggland}, L., {Hansteen}, V.~H., {De Pontieu}, B., \& {Carlsson}, M. 2011,
  \apj, 743, 142, \dodoi{10.1088/0004-637X/743/2/142}

\bibitem[{{Hindman} \& {Jain}(2008)}]{hindman_jain_2008}
{Hindman}, B.~W., \& {Jain}, R. 2008, \apj, 677, 769, \dodoi{10.1086/528956}

\bibitem[{{Jefferies} {et~al.}(2006){Jefferies}, {McIntosh}, {Armstrong},
  {Bogdan}, {Cacciani}, \& {Fleck}}]{jefferies_etal_2006}
{Jefferies}, S.~M., {McIntosh}, S.~W., {Armstrong}, J.~D., {et~al.} 2006,
  \apjl, 648, L151, \dodoi{10.1086/508165}

\bibitem[{{Jess} {et~al.}(2020){Jess}, {Snow}, {Houston}, {Botha}, {Fleck},
  {Krishna Prasad}, {Asensio Ramos}, {Morton}, {Keys}, {Jafarzadeh},
  {Stangalini}, {Grant}, \& {Christian}}]{jess_etal_2020}
{Jess}, D.~B., {Snow}, B., {Houston}, S.~J., {et~al.} 2020, Nature Astronomy,
  4, 220, \dodoi{10.1038/s41550-019-0945-2}

\bibitem[{{Karampelas} \& {Van
  Doorsselaere}(2021)}]{karampelas_van_doorsselaere_2021}
{Karampelas}, K., \& {Van Doorsselaere}, T. 2021, arXiv e-prints,
  arXiv:2102.03332.
\newblock \doarXiv{2102.03332}

\bibitem[{{Kerner} {et~al.}(1998){Kerner}, {Goedbloed}, {Huysmans}, {Poedts},
  \& {Schwarz}}]{kerner_etal_1998}
{Kerner}, W., {Goedbloed}, J.~P., {Huysmans}, G.~T.~A., {Poedts}, S., \&
  {Schwarz}, E. 1998, Journal of Computational Physics, 142, 271,
  \dodoi{10.1006/jcph.1998.5910}

\bibitem[{{Kerner} {et~al.}(1985){Kerner}, {Lerbinger}, {Gruber}, \&
  {Tsunematsu}}]{kerner_etal_1985}
{Kerner}, W., {Lerbinger}, K., {Gruber}, R., \& {Tsunematsu}, T. 1985, Computer
  Physics Communications, 36, 225, \dodoi{10.1016/0010-4655(85)90053-0}

\bibitem[{{Keys} {et~al.}(2018){Keys}, {Morton}, {Jess}, {Verth}, {Grant},
  {Mathioudakis}, {Mackay}, {Doyle}, {Christian}, {Keenan}, \&
  {Erd{\'e}lyi}}]{keys_etal_2018}
{Keys}, P.~H., {Morton}, R.~J., {Jess}, D.~B., {et~al.} 2018, \apj, 857, 28,
  \dodoi{10.3847/1538-4357/aab432}

\bibitem[{{Khomenko} \& {Cally}(2012)}]{khomenko+cally_2012}
{Khomenko}, E., \& {Cally}, P.~S. 2012, \apj, 746, 68,
  \dodoi{10.1088/0004-637X/746/1/68}

\bibitem[{{Khomenko} \& {Cally}(2019)}]{khomenko_cally_2019_part_II}
---. 2019, \apj, 883, 179, \dodoi{10.3847/1538-4357/ab3d28}

\bibitem[{{Khomenko} {et~al.}(2008){Khomenko}, {Centeno}, {Collados}, \&
  {Trujillo Bueno}}]{khomenko_etal_2008_letter}
{Khomenko}, E., {Centeno}, R., {Collados}, M., \& {Trujillo Bueno}, J. 2008,
  \apjl, 676, L85, \dodoi{10.1086/587057}

\bibitem[{{Khomenko} \& {Collados}(2006)}]{khomenko_collados_2006}
{Khomenko}, E., \& {Collados}, M. 2006, \apj, 653, 739, \dodoi{10.1086/507760}

\bibitem[{{Khomenko} \& {Collados}(2015)}]{khomenko_collados_2015_review}
---. 2015, Living Reviews in Solar Physics, 12, 6, \dodoi{10.1007/lrsp-2015-6}

\bibitem[{Lamb(1909)}]{lamb_1909}
Lamb, H. 1909, Proceedings of the London Mathematical Society, s2-7, 122,
  \dodoi{10.1112/plms/s2-7.1.122}

\bibitem[{{Leighton}(1960)}]{leighton_1960}
{Leighton}, R.~B. 1960, in Aerodynamic Phenomena in Stellar Atmospheres, ed.
  R.~N. {Thomas}, Vol.~12, 321--325

\bibitem[{{Liu} {et~al.}(2011){Liu}, {Title}, {Zhao}, {Ofman}, {Schrijver},
  {Aschwanden}, {De Pontieu}, \& {Tarbell}}]{liu_etal_2011}
{Liu}, W., {Title}, A.~M., {Zhao}, J., {et~al.} 2011, \apjl, 736, L13,
  \dodoi{10.1088/2041-8205/736/1/L13}

\bibitem[{{McIntosh} \& {Jefferies}(2006)}]{mcintosh+jefferies_2006}
{McIntosh}, S.~W., \& {Jefferies}, S.~M. 2006, \apjl, 647, L77,
  \dodoi{10.1086/507425}

\bibitem[{{Mignone} {et~al.}(2007){Mignone}, {Bodo}, {Massaglia}, {Matsakos},
  {Tesileanu}, {Zanni}, \& {Ferrari}}]{mignone_etal_2007}
{Mignone}, A., {Bodo}, G., {Massaglia}, S., {et~al.} 2007, \apjs, 170, 228,
  \dodoi{10.1086/513316}

\bibitem[{{Mignone} {et~al.}(2018){Mignone}, {Bodo}, {Vaidya}, \&
  {Mattia}}]{mignone_etal_2018}
{Mignone}, A., {Bodo}, G., {Vaidya}, B., \& {Mattia}, G. 2018, \apj, 859, 13,
  \dodoi{10.3847/1538-4357/aabccd}

\bibitem[{{Mignone} {et~al.}(2012){Mignone}, {Zanni}, {Tzeferacos}, {van
  Straalen}, {Colella}, \& {Bodo}}]{mignone_etal_2012}
{Mignone}, A., {Zanni}, C., {Tzeferacos}, P., {et~al.} 2012, \apjs, 198, 7,
  \dodoi{10.1088/0067-0049/198/1/7}

\bibitem[{{Mihalas} \& {Mihalas}(1984)}]{mihalas_mihalas_1984}
{Mihalas}, D., \& {Mihalas}, B.~W. 1984, {Foundations of radiation
  hydrodynamics} (Oxford: Oxford university press)

\bibitem[{{Moreels} {et~al.}(2013){Moreels}, {Goossens}, \& {Van
  Doorsselaere}}]{moreels_etal_2013}
{Moreels}, M.~G., {Goossens}, M., \& {Van Doorsselaere}, T. 2013, \aap, 555,
  A75, \dodoi{10.1051/0004-6361/201321545}

\bibitem[{{Moreels} \& {Van Doorsselaere}(2013)}]{moreels_vandoorsselaere_2013}
{Moreels}, M.~G., \& {Van Doorsselaere}, T. 2013, \aap, 551, A137,
  \dodoi{10.1051/0004-6361/201219568}

\bibitem[{{Moreels} {et~al.}(2015){Moreels}, {Van Doorsselaere}, {Grant},
  {Jess}, \& {Goossens}}]{moreels_etal_2015}
{Moreels}, M.~G., {Van Doorsselaere}, T., {Grant}, S.~D.~T., {Jess}, D.~B., \&
  {Goossens}, M. 2015, \aap, 578, A60, \dodoi{10.1051/0004-6361/201425468}

\bibitem[{{Morton} {et~al.}(2011){Morton}, {Erd{\'e}lyi}, {Jess}, \&
  {Mathioudakis}}]{morton_etal_2011}
{Morton}, R.~J., {Erd{\'e}lyi}, R., {Jess}, D.~B., \& {Mathioudakis}, M. 2011,
  \apjl, 729, L18, \dodoi{10.1088/2041-8205/729/2/L18}

\bibitem[{{Morton} {et~al.}(2016){Morton}, {Tomczyk}, \&
  {Pinto}}]{morton_etal_2016}
{Morton}, R.~J., {Tomczyk}, S., \& {Pinto}, R.~F. 2016, \apj, 828, 89,
  \dodoi{10.3847/0004-637X/828/2/89}

\bibitem[{{Morton} {et~al.}(2012){Morton}, {Verth}, {Jess}, {Kuridze},
  {Ruderman}, {Mathioudakis}, \& {Erd{\'e}lyi}}]{morton_etal_2012}
{Morton}, R.~J., {Verth}, G., {Jess}, D.~B., {et~al.} 2012, Nature
  Communications, 3, 1315, \dodoi{10.1038/ncomms2324}

\bibitem[{{Morton} {et~al.}(2019){Morton}, {Weberg}, \&
  {McLaughlin}}]{morton_etal_2019}
{Morton}, R.~J., {Weberg}, M.~J., \& {McLaughlin}, J.~A. 2019, Nature
  Astronomy, \dodoi{10.1038/s41550-018-0668-9}

\bibitem[{{Nakariakov} {et~al.}(2016){Nakariakov}, {Anfinogentov},
  {Nistic{\`o}}, \& {Lee}}]{nakariakov_etal_2016}
{Nakariakov}, V.~M., {Anfinogentov}, S.~A., {Nistic{\`o}}, G., \& {Lee}, D.-H.
  2016, \aap, 591, L5, \dodoi{10.1051/0004-6361/201628850}

\bibitem[{{Nakariakov} \& {Kolotkov}(2020)}]{nakariakov_kolotkov_2020_review}
{Nakariakov}, V.~M., \& {Kolotkov}, D.~Y. 2020, \araa, 58, 441,
  \dodoi{10.1146/annurev-astro-032320-042940}

\bibitem[{{Nakariakov} \& {Verwichte}(2005)}]{nakariakov_verwichte_2005_review}
{Nakariakov}, V.~M., \& {Verwichte}, E. 2005, Living Reviews in Solar Physics,
  2, 3, \dodoi{10.12942/lrsp-2005-3}

\bibitem[{{Nistic{\`o}} {et~al.}(2013){Nistic{\`o}}, {Nakariakov}, \&
  {Verwichte}}]{nistico_etal_2013}
{Nistic{\`o}}, G., {Nakariakov}, V.~M., \& {Verwichte}, E. 2013, \aap, 552,
  A57, \dodoi{10.1051/0004-6361/201220676}

\bibitem[{Powell {et~al.}(1999)Powell, Roe, Linde, Gombosi, \& {De
  Zeeuw}}]{powell_etal_1999}
Powell, K.~G., Roe, P.~L., Linde, T.~J., Gombosi, T.~I., \& {De Zeeuw}, D.~L.
  1999, Journal of Computational Physics, 154, 284 ,
  \dodoi{https://doi.org/10.1006/jcph.1999.6299}

\bibitem[{Priest(2014)}]{priest_2014}
Priest, E. 2014, Magnetohydrodynamics of the Sun (Cambridge University Press),
  \dodoi{10.1017/CBO9781139020732}

\bibitem[{{Rajaguru} {et~al.}(2019){Rajaguru}, {Sangeetha}, \&
  {Tripathi}}]{rajaguru_etal_2019}
{Rajaguru}, S.~P., {Sangeetha}, C.~R., \& {Tripathi}, D. 2019, \apj, 871, 155,
  \dodoi{10.3847/1538-4357/aaf883}

\bibitem[{{Reale} {et~al.}(2016){Reale}, {Orlando}, {Guarrasi}, {Mignone},
  {Peres}, {Hood}, \& {Priest}}]{reale_etal_2016}
{Reale}, F., {Orlando}, S., {Guarrasi}, M., {et~al.} 2016, \apj, 830, 21,
  \dodoi{10.3847/0004-637X/830/1/21}

\bibitem[{{Riedl} {et~al.}(2021){Riedl}, {Gilchrist-Millar}, {Van
  Doorsselaere}, {Jess}, \& {Grant}}]{riedl_etal_2021}
{Riedl}, J.~M., {Gilchrist-Millar}, C.~A., {Van Doorsselaere}, T., {Jess},
  D.~B., \& {Grant}, S.~D.~T. 2021, \aap, 648, A77,
  \dodoi{10.1051/0004-6361/202040163}

\bibitem[{{Riedl} {et~al.}(2019){Riedl}, {Van Doorsselaere}, \&
  {Santamaria}}]{riedl_etal_2019}
{Riedl}, J.~M., {Van Doorsselaere}, T., \& {Santamaria}, I.~C. 2019, \aap, 625,
  A144, \dodoi{10.1051/0004-6361/201935393}

\bibitem[{{Roberts}(2006)}]{roberts_2006}
{Roberts}, B. 2006, Philosophical Transactions of the Royal Society of London
  Series A, 364, 447, \dodoi{10.1098/rsta.2005.1709}

\bibitem[{{Roberts}(2019)}]{roberts_2019_book}
---. 2019, MHD waves in the solar atmosphere (Cambridge: Cambridge University
  Press)

\bibitem[{{Santamaria} {et~al.}(2015){Santamaria}, {Khomenko}, \&
  {Collados}}]{santamaria_etal_2015}
{Santamaria}, I.~C., {Khomenko}, E., \& {Collados}, M. 2015, \aap, 577, A70,
  \dodoi{10.1051/0004-6361/201424701}

\bibitem[{{Schmitz} \& {Fleck}(1998)}]{schmitz_fleck_1998}
{Schmitz}, F., \& {Fleck}, B. 1998, \aap, 337, 487

\bibitem[{{Schunker} \& {Cally}(2006)}]{schunker_cally_2006}
{Schunker}, H., \& {Cally}, P.~S. 2006, \mnras, 372, 551,
  \dodoi{10.1111/j.1365-2966.2006.10855.x}

\bibitem[{{Soler} {et~al.}(2013){Soler}, {Goossens}, {Terradas}, \&
  {Oliver}}]{soler_etal_2013}
{Soler}, R., {Goossens}, M., {Terradas}, J., \& {Oliver}, R. 2013, \apj, 777,
  158, \dodoi{10.1088/0004-637X/777/2/158}

\bibitem[{{Srivastava} {et~al.}(2021){Srivastava}, {Ballester}, {Cally},
  {Carlsson}, {Goossens}, {Jess}, {Khomenko}, {Mathioudakis}, {Murawski}, \&
  {Zaqarashvili}}]{srivastava_etal_2021_review}
{Srivastava}, A.~K., {Ballester}, J.~L., {Cally}, P.~S., {et~al.} 2021, Journal
  of Geophysical Research: Space Physics, 126, e2020JA029097,
  \dodoi{https://doi.org/10.1029/2020JA029097}

\bibitem[{{Stenuit} {et~al.}(1998){Stenuit}, {Keppens}, \&
  {Goossens}}]{stenuit_etal_1998}
{Stenuit}, H., {Keppens}, R., \& {Goossens}, M. 1998, \aap, 331, 392

\bibitem[{{Stenuit} {et~al.}(1999){Stenuit}, {Tirry}, {Keppens}, \&
  {Goossens}}]{stenuit_etal_1999}
{Stenuit}, H., {Tirry}, W.~J., {Keppens}, R., \& {Goossens}, M. 1999, \aap,
  342, 863

\bibitem[{{Strang}(1968)}]{strang_1968}
{Strang}, G. 1968, SIAM Journal on Numerical Analysis, 5, 506,
  \dodoi{10.1137/0705041}

\bibitem[{{Tarr} {et~al.}(2017){Tarr}, {Linton}, \& {Leake}}]{tarr_etal_2017}
{Tarr}, L.~A., {Linton}, M., \& {Leake}, J. 2017, \apj, 837, 94,
  \dodoi{10.3847/1538-4357/aa5e4e}

\bibitem[{{Tomczyk} \& {McIntosh}(2009)}]{tomczyk_mcintosh_2009}
{Tomczyk}, S., \& {McIntosh}, S.~W. 2009, \apj, 697, 1384,
  \dodoi{10.1088/0004-637X/697/2/1384}

\bibitem[{{Tomczyk} {et~al.}(2007){Tomczyk}, {McIntosh}, {Keil}, {Judge},
  {Schad}, {Seeley}, \& {Edmondson}}]{tomczyk_etal_2007}
{Tomczyk}, S., {McIntosh}, S.~W., {Keil}, S.~L., {et~al.} 2007, Science, 317,
  1192, \dodoi{10.1126/science.1143304}

\bibitem[{{Van Doorsselaere} {et~al.}(2008){Van Doorsselaere}, {Nakariakov},
  {Young}, \& {Verwichte}}]{van_doorsselaere_etal_2008_b}
{Van Doorsselaere}, T., {Nakariakov}, V.~M., {Young}, P.~R., \& {Verwichte}, E.
  2008, \aap, 487, L17, \dodoi{10.1051/0004-6361:200810186}

\bibitem[{{Van Doorsselaere} \& {Poedts}(2007)}]{van_doorsselaere+poedts_2007}
{Van Doorsselaere}, T., \& {Poedts}, S. 2007, Plasma Physics and Controlled
  Fusion, 49, 261, \dodoi{10.1088/0741-3335/49/3/006}

\bibitem[{{Van Doorsselaere} {et~al.}(2020){Van Doorsselaere}, {Srivastava},
  {Antolin}, {Magyar}, {Vasheghani Farahani}, {Tian}, {Kolotkov}, {Ofman},
  {Guo}, {Arregui}, {De Moortel}, \&
  {Pascoe}}]{vandoorsselaere_etal_2020_review}
{Van Doorsselaere}, T., {Srivastava}, A.~K., {Antolin}, P., {et~al.} 2020,
  \ssr, 216, 140, \dodoi{10.1007/s11214-020-00770-y}

\bibitem[{{Verwichte} {et~al.}(2005){Verwichte}, {Nakariakov}, \&
  {Cooper}}]{verwichte_etal_2005}
{Verwichte}, E., {Nakariakov}, V.~M., \& {Cooper}, F.~C. 2005, \aap, 430, L65,
  \dodoi{10.1051/0004-6361:200400133}

\bibitem[{{Wilson}(1981)}]{wilson_1981}
{Wilson}, P.~R. 1981, \apj, 251, 756, \dodoi{10.1086/159519}

\bibitem[{{Zhang} {et~al.}(2020){Zhang}, {Dai}, {Xu}, {Li}, {Lu}, {Tam}, \&
  {Xu}}]{zhang_etal_2020}
{Zhang}, Q.~M., {Dai}, J., {Xu}, Z., {et~al.} 2020, A\&A, 638, A32,
  \dodoi{10.1051/0004-6361/202038233}

\bibitem[{{Zhugzhda} \& {Locans}(1981)}]{zhugzhda+locans_1981}
{Zhugzhda}, Y.~D., \& {Locans}, V. 1981, Soviet Astronomy Letters, 7, 25

\bibitem[{{Zurbriggen} {et~al.}(2020){Zurbriggen}, {Sieyra}, {Costa},
  {Esquivel}, \& {Stenborg}}]{zurbriggen_etal_2020}
{Zurbriggen}, E., {Sieyra}, M.~V., {Costa}, A., {Esquivel}, A., \& {Stenborg},
  G. 2020, \mnras, 494, 5270, \dodoi{10.1093/mnras/staa1105}

\end{thebibliography}
\bibliographystyle{aasjournal}



\end{document}